\documentclass[aip,reprint,cha,amsmath,amsfonts,amssymb]{revtex4-1}
\usepackage{here}
\usepackage{docs}%
\usepackage{graphicx}
\usepackage{amsmath,amsfonts,amssymb}
\usepackage{color}
\usepackage{bm}
\usepackage{ascmac}
\usepackage{soul,xcolor}

\usepackage{threeparttable}
\usepackage[mathscr]{eucal}
\usepackage{titlesec}
\usepackage{picture}
\usepackage{caption}

\expandafter\ifx\csname package@font\endcsname\relax\else
 \expandafter\expandafter
 \expandafter\usepackage
 \expandafter\expandafter
 \expandafter{\csname package@font\endcsname}%
\fi
\usepackage{titlesec}
\usepackage{picture}
\definecolor{teal}{RGB}{0,128,128}
\definecolor{powderblue}{RGB}{176,224,230}
\definecolor{darkslateblue}{RGB}{72,61,139}
\definecolor{darkslategray}{RGB}{47,79,79}
\definecolor{lightdarkslateblue}{RGB}{224,255,255}
\definecolor{gray}{rgb}{0.6,0.6,0.6}
\setstcolor{purple}
\allowdisplaybreaks[4]

\newcommand{\bra}{\langle}
\newcommand{\ket}{\rangle}
\newcommand{\dd}{\partial}
\newcommand{\bre}{\nonumber\\}

\definecolor{green1}{rgb}{0.2,0.6,0.2}
\definecolor{orange}{RGB}{255,165,0}
\definecolor{yellow}{RGB}{255,241,15}

\definecolor{dblue}{RGB}{50,50,170}

\definecolor{dgren}{RGB}{16,95,55}

\definecolor{brown}{RGB}{111,16,50}

\newcommand{\black}{\color{black}}
\definecolor{purple}{rgb}{0.8,0.0,0.8}
\definecolor{pink}{rgb}{1.,0.5,0.5}

\newcommand{\Roo}{{\bf R}^{\rm oo}_g}
\newcommand{\Rvo}{{\bf R}_g^{\rm vo}}
\newcommand{\Rov}{{\bf R}_g^{\rm ov}}
\newcommand{\Rvv}{{\bf R}_g^{\rm vv}}
\newcommand{\Woo}{{\bm {\mathcal W}}_{g}^{\rm oo}}
\newcommand{\Wvo}{{\bm {\mathcal W}}_{g}^{\rm vo}}

\newcommand{\Wvv}{{\bm {\mathcal W}}_{g}^{\rm vv}}
\newcommand{\bC}{{\bf C}}

\newcommand{\bS}{{\bf S}}
\newcommand{\Rg}{{\bf R}_g}
\newcommand{\PP}{{\mathscr P}}

\newcommand{\calE}{{\mathcal E}}
\newcommand{\calN}{{\mathcal N}}
\newcommand{\Fg}{{\bf F}_g}

\newcommand{\Lg}{{\bm {\mathcal L}}_g}
\newcommand{\R}{{\bm {\mathcal R}}_g}
\newcommand{\pg}{{\bm \rho}_g}

\newcommand{\CSC}{\bC^\dag \bS^x \bC}
\newcommand{\calW}{{\mathcal W}}
\newcommand{\Wg}{{\bm{\mathcal W}}_g}
\newcommand{\calL}{{\mathcal L}}
\newcommand{\calR}{{\mathcal R}}
\newcommand{\calF}{{\mathcal F}}

\newcommand{\Tr}{{\rm Tr}}
\newcommand{\tcalF}{\tilde {\mathcal F}}
\newcommand{\calV}{\bar {\mathcal V}}

\begin{document}

\title{General technique for analytical derivatives of post-projected Hartree-Fock}
\author{Takashi Tsuchimochi}
\email{tsuchimochi@gmail.com}
\affiliation{Graduate School of Science, Technology, and Innovation, Kobe University, Kobe, Hyogo 657-0025 Japan}
\author{Seiichiro Ten-no}
\email{tenno@cs.kobe-u.ac.jp}
\affiliation{Graduate School of Science, Technology, and Innovation, Kobe University, Kobe, Hyogo 657-0025 Japan}
\affiliation{Graduate School of System Informatics, Kobe University, Kobe, Hyogo 657-0025 Japan}%\date{\today}
\begin{abstract}
In electronic structure theory, the availability of analytical derivative is one of the desired features for a method to be useful in practical applications, as it allows for geometry optimization as well as computation of molecular properties. With the recent advances in the development of symmetry-projected Hartree-Fock (PHF) methods, we here aim at further extensions by devising the analytic gradients of post-PHF approaches with a special focus on spin-extended (spin-projected) configuration interaction with single and double substitutions (ECISD). Just like standard single-reference methods, the mean-field PHF part does not require the corresponding coupled-perturbed equation to be solved, while the correlation energy term needs the orbital relaxation effect to be accounted for, unless the underlying molecular orbitals are variationally optimized in the presence of the correlation energy. We present a general strategy for post-PHF analytical gradients, which closely parallels that for single-reference methods, yet addressing the major difference between them. The similarity between ECISD and multi-reference CI not only in the energy but also in the optimized geometry is clearly demonstrated by the numerical examples of ozone and cyclobutadiene.

\end{abstract}
\maketitle
\section{Introduction}
Recently, we have proposed and developed novel wave function methods based on symmetry-projected Hartree-Fock\cite{Lowdin55A,Lowdin55B,Scuseria11,Jimenez12} (PHF) in analogy with the traditional single reference approaches.\cite{Tsuchimochi14,Tsuchimochi15A,Tsuchimochi15C,Tsuchimochi16A,Tsuchimochi16B} It appears PHF successfully treats the major static (nondynamical) correlation effect arising due to electronic degeneracies in a system, by breaking and restoring the symmetries that a wave function ought to possess.   The key feature of symmetry-breaking is that a broken-symmetry  HF determinant such as spin-unrestricted HF (UHF) can be written as a linear combination of {\it multiple} configuration state functions, each of which is highly multi-reference (MR) possessing a proper symmetry. 
As a result, such an ``effectively MR'' picture hidden in a broken-symmetry Slater determinant is potentially capable of providing an efficient means to account for static correlation.\cite{Tsuchimochi09,Scuseria09}
A symmetry projection operator then eliminates the undesired components with the irrelevant symmetries, retrieving a \emph{genuine} MR wave function with the designated symmetry.
Hence, PHF stands as an improved alternative to RHF and UHF, while being black-box unlike most traditional MR methods.
This fact has motivated us to develop post-PHF methods to account for the residual dynamical correlation by applying perturbation theory\cite{Tsuchimochi14} (EMP2) or configuration interaction\cite{Tsuchimochi16A,Tsuchimochi16B} (ECI), without necessiating very expensive canonical MR calculations.
Combining PHF with density functional correlations was also shown to improve the quantitative accuracy by Garza and coworkers.\cite{Garza13,Garza14}
Another direction of post-PHF developments includes a time-dependent extension\cite{Tsuchimochi15A} (TDPHF) and introduction of spin-flip (SF) excitations\cite{Krylov01A,Krylov01B,Krylov06,Tsuchimochi15C} for excited states.
However, most benchmark calculations in these works have focused only on energetics, and molecular properties such as dipole moment have been hardly studied.
This is due to the fact that all these post-PHF methods do not satisfy the Hellmann-Feynman theorem, despite they have well-defined wave functions or energy functionals.

To account for the non-Hellmann-Feynman contribution, one needs to find the energy derivative with respect to an infinitesimal perturbation.
To see this, let us introduce a one-electron perturbation $\lambda\hat O =  \sum_{\mu\nu} \bra \phi_\mu |\lambda\hat O| \phi_{\nu}\ket a_\mu^\dag a_\nu$ into the Hamiltonian $\hat H$,
\begin{align}
\hat H(\lambda) = \hat H + \lambda \hat O,
\end{align}
where we assume basis functions do not depend on $\lambda$ for the sake of simplicity.
The Taylor expansion of $\calE(\lambda) = \bra \Psi|\hat H(\lambda) |\Psi\ket$ around $\lambda=0$ becomes
\begin{align}
\calE(\lambda) = \calE(0) + \lambda \left.\frac{d\calE}{d\lambda}\right|_{\lambda=0} + \frac{1}{2}\lambda^2 \left.\frac{d^2\calE}{d\lambda^2}\right|_{\lambda=0} + \cdots.
\end{align} 
Since approximate wave functions do not in general satisfy the Hellmann-Feynman theorem, the expectation value of $\hat O$ is different from the first-order energy derivative.
It is widely accepted that the derivative approach is more appropriate because it accounts for the response of perturbation. The one-electron property with $\hat O$ can be computed by
\begin{align}
\left.\frac{d \calE(\lambda)}{d\lambda}\right|_{\lambda = 0} = \sum_{\mu\nu}  \bra \phi_\mu |\hat O| \phi_{\nu}\ket D_{\nu\mu}^{\rm rel},
\end{align}
where ${\bf D}^{\rm rel}$ is the relaxed density. The molecular properties computed without the appropriate response correction are known to be inaccurate for approximate methods, and the importance of the non-Hellmann-Feynman contribution in calculation of molecular properties is well documented.\cite{Diercksen81,Almlof85}%For example, the dipole moment of ozone computed by ECI with singles and doubles (ECISD) is 0.274 $D$ with cc-PVTZ, largely underestimating the experimental value 0.532 $D$. 

PHF does not suffer from this problem since it satisfies the Hellmann-Feynman theorem if the $\lambda$ dependence is not present in basis functions. If, on the other hand, $\lambda$ is a nuclear displacement $x$, then PHF also requires the non-Hellmann-Feynman term, the so-called Pulay force.\cite{Pulay69,Pulay70} Yet, the fully variational nature of the PHF wave function makes the derivation straightforward without the necessity of solving the coupled-perturbed (CP) equations\cite{Gerratt68} or equivalently the Z-vector equation\cite{Handy84} for PHF.
In fact, the analytical nuclear derivative of PHF was first derived by Schutski {\it et al.} except for the point-group symmetry projection,\cite{Schutski14} which was later incorporated by Uejima and Ten-no.\cite{Uejima16}
The latter authors also derived the nuclear gradient for the projection-after-variation scheme, where the orbitals used are variationally determined by UHF, and therefore the corresponding CPHF equation was required in general with less computational cost for the self-consistent field (SCF) part.
  
Taking these works into consideration, our next step is to devise the analytical gradients of post-PHF methods with respect to nuclear displacements, because not only is it straightforward to obtain the relaxed density matrix ${\bf D}^{\rm rel}$ from the resulting expressions, but also it will enable us to perform geometry optimization. In this paper, therefore we will give a detailed discussion on the analytical derivative technique with a special focus on the recently developed ECI with singles and doubles (ECISD)\cite{Tsuchimochi16A,Tsuchimochi16B} as an illustrative example. Hence, the projection operator we consider is of collinear spin-symmetry, i,e, the reference is spin-projected UHF (SUHF); this is reasonable as the most post-PHF schemes developed so far are based on collinear spin-projection. Nonetheless, we first give a general perspective for post-PHF gradients, including other symmetry-projections, in order to point out that it is straightforward to generalize the working equations of ECISD although the resulting derivations would be necessarily cumbersome. Also the strategy outlined below is applicable to any other levels of theory than ECISD; for example, geometry optimization of TDPHF excited states are possible. Their specific derivations and performances will be discussed in the forthcoming papers.

This paper is organized as follows. In Section \ref{sec:General}, we will outline the general workflow for analytical derivative of post-PHF methods, with some notes on the difference from post-HF ones. SUHF, the nonorthogonal Wick theorem, and ECISD are briefly reviewed in Section \ref{sec:Nomen} to define our notations used throughout this paper. Section \ref{sec:ECISDFock} and \ref{sec:CPPHF} provide the working equations for the ECISD orbital gradient and CPPHF equations. Section \ref{sec:Explicit} deals with the explicit dependence of the energy on the nuclear displacement. A short remark on the size-consistent correction to ECISD is given in Section \ref{sec:ECISD+Q}. Finally, we demonstrate the validity and performance of ECISD energy gradients for ozone and cyclobutadiene.

\section{Theory}\label{sec:Theory}
\subsection{General view for analytical derivative of post-PHF methods}\label{sec:General}
We first consider the analytical nuclear gradients of post-PHF methods whose total energy $\calE$ may be variational with respect to particle-hole amplitudes $c_I$, e.g., CI coefficients.
Due to the restriction that MO coefficients {\bf C} are optimal with PHF energy $E_{\rm PHF}$ but not with $\calE$ itself, one has to take into account the effect of the possible orbital change upon a small perturbation $x$, either explicitly or implicitly. One way is to formulate and solve linear equations to compute $\frac{\dd {\bf C}}{\dd x}$ for $3N_{\rm atom}$ times, just like CPHF.
For traditional post-HF methods, Handy and Schaefer have shown that these linear equations can be simplified to only one set of equations, using their Z-vector method.\cite{Handy84} A different formulation of the Z-vector method was realized by Helgaker {et al.},\cite{Helgaker89,Jorgensen88} which uses the Lagrangian multiplier approach and has later gained popularity in electronic structure theory because it has extended the applicability of the Z-vector method to any order of energy derivatives.\cite{Hald03,Celani03,Kaduk14,Uejima16} We will closely follow Helgaker's scheme as it is intuitive, simple, and equally applicable to PHF based methods.

In our problem, {\bf C} is determined by minimizing $E_{\rm PHF}$ instead of $E_{\rm HF}$ under the condition that orbitals are mutually orthonormal. This means the generalized Brillouin theorem holds; given the PHF energy
\begin{align}
E_{\rm PHF} = \frac{\bra \Xi_{\cal J, \cal M} | \hat H |\Xi_{\cal J,\cal M}\ket}{ \bra \Xi_{\cal J, \cal M} | \Xi_{\cal J,\cal M}\ket},
\end{align}  
with the corresponding PHF state
\begin{align}
|\Xi_{\cal J, \cal M} \ket =\sum_{\cal K} f_{\cal K} \hat {\cal P}^{\cal J}_{\cal M \cal K}|\Phi\ket \label{eq:generalP},
\end{align}
the PHF Fock matrix elements are all zero at the self-consistency of  a PHF state $|\Xi_{\cal J,\cal M}\ket$:
\begin{align}
\calF_{ai} = \frac{\sum_{\cal K \cal K'} f_{\cal K}^* f_{\cal K'}  \bra \Phi| a^\dag_i a_a (\hat H - E_{\rm PHF} ) \hat {\cal P}^{\cal J}_{\cal K K'} |\Phi\ket}{ \sum_{\cal K\cal K'} f^*_{\cal K} f_{\cal K'} \bra \Phi | \hat {\cal P}^{\cal J}_{\cal K K'} |\Phi\ket} = 0,\label{eq:Fai}
%\calF_{ia} = \frac{ \bra \Phi| (\hat H - E_{\rm PHF} ) \hat P a^\dag_a a_i |\Phi\ket} { \bra \Phi | \hat P |\Phi\ket},
\end{align}
where we have used the fact that $\hat {\cal P}_{\cal MK}^{\cal J}$ is commutable with $\hat H$ and is orthogonal to each other.\cite{Jimenez12}
Here $|\Phi\ket$ is a broken-symmetry HF determinant and $\cal J, K, \cdots$ represent different symmetry components.  Also, throughout this paper, we will use the {\it spin}-orbitals unless otherwise noted, with $i,j,...$ to indicate occupied orbitals, $a,b,...$ virtual orbitals, $p,q,...$ general orbitals, and $\mu,\nu,...$ atomic orbitals (AO). Therefore, generally, each index runs over both $\alpha$ and $\beta$ spins. We should mention that, in our notation, a matrix may be represented in either AO or MO, depending on the characters used for indices.

In addition to the orbital condition Eq.(\ref{eq:Fai}), $f_{\cal K}$ are also variationally determined by diagonalizing the Hamiltonian matrix ${\cal H}_{\cal KK'}=\bra \Phi| \hat H \hat {\cal P}_{\cal K \cal K'} |\Phi\ket$ under metric $\calN_{\cal KK'}=\bra \Phi| \hat {\cal P}_{\cal K \cal K'} |\Phi\ket$. %Now, the condition that Eq.(\ref{eq:Fai}) is zero is always satisfied even when perturbation $x$ is applied to the system. 
Hence, similarly to many post-HF methods, we set up the following Lagrangian,
\begin{align}
{\mathscr L} =& \calE[{\bf c},{\bf C},{\bf f}]+ \sum_{ia} \left( z_{ai} {\mathcal F}_{ai} + z_{ia} {\mathcal F}_{ia} \right) \bre
&+ \sum_{\cal KK'}\Bigl[ {\cal Z}_{\cal K} \left({\cal H}_{\cal KK'} - E_{\rm PHF} {\cal N}_{\cal KK'}\right)f_{\cal K'}  \bre 
& - \Lambda \left({\cal N}_{\cal KK'} f_{\cal K}^* f_{\cal K'}  -1 \right)\Bigr]
 - \Tr [{\bm \epsilon}\left(\bC^\dag \bS \bC - {\bf 1}\right)],\label{eq:Lag}
\end{align} 
where ${\bf z} = {\bf z}^\dag$ and ${\bm{\mathcal Z}}$ are the Lagrange multipliers to account for the first-order relaxation effect of {\bf C} and {\bf f} in the presence of correlation, and therefore will play a role of response. Note that nonzero {\bf z} and ${\bm{\mathcal Z}}$ are due to the absence of the Hellmann-Feynman theorem in $\calE$. The dimension of {\bf z} changes depending on the PHF scheme employed; for non-collinear PHF, the $\alpha\beta$ and $\beta\alpha$ components are nonzero, while for collinear SUHF, they are redundant and necessarily zero. The terms with $\Lambda$ and $\bm\epsilon$ in Eq.(\ref{eq:Lag}) are added due to the orthonormal condition of ${\bf f}$ and ${\bf C}$ under metric ${\bm{\mathcal N}}$ and  AO overlap matrix {\bf S}. Finally, note that the normalization condition of {\bf c} is not explicitly treated here for the sake of simplicity, but it is implicitly included in the denominator of $\cal E$.

With parameters ${\bf c}$ determined variationally, i.e., $\frac{\dd \calE}{\dd {\bf c}} = 0$, we find 
\begin{align}
\frac{\dd {\mathscr L}}{\dd c_I} = \frac{\dd {\mathscr L}}{\dd \epsilon_{pq}} =  \frac{\dd {\mathscr L}}{\dd z_{ia}}  = \frac{\dd {\mathscr L}}{\dd {\cal Z}_{\cal K}} =
\frac{\dd {\mathscr L}}{\dd\Lambda} = 0,
\end{align}
and therefore using the chain-rule
\begin{align}
\frac{d \calE}{d x} \equiv \frac{d {\mathscr L}}{d x} 
%&= \frac{\dd {\mathscr L}}{\dd x} + \frac{\dd {\mathscr L}}{\dd {\bf c}}\frac{d {\bf c}}{d x} + \frac{\dd {\mathscr L}}{\dd {\bf z}}\frac{d{\bf z}}{dx} + \frac{\dd {\mathscr L}}{\dd {\bm\epsilon}}\frac{d{\bm\epsilon}}{dx} + \frac{\dd {\mathscr L}}{\dd {\bf C}}\frac{d {\bf C}}{d x} \bre
&=  \frac{\dd {\mathscr L}}{\dd x} + \frac{\dd {\mathscr L}}{\dd {\bf C}}\frac{d {\bf C}}{d x} + \frac{\dd {\mathscr L}}{\dd {\bf f}}\frac{d {\bf f}}{d x}.
\end{align}
The troublesome $\frac{d {\bf C}}{d x}$ and $\frac{d {\bf f}}{d x}$ terms shall not enter the equation if we enforce $\frac{\dd {\mathscr L}}{\dd {\bf C}} = \frac{\dd {\mathscr L}}{\dd {\bf f}} = {\bf 0}$. 

Below, instead of using {\bf C} itself, we parametrize {\bf C} as
\begin{align}
{\bf C}[{\bm\kappa}] = {\bf C}_0 \exp({\bm\kappa}),
\end{align}
for convenience, where ${\bm\kappa}$ is an anti-Hermitian matrix and $\exp({\bm\kappa})$ performs an orbital rotation from the reference ${\bf C}_0$, which is set to constant. Then our task boils down to  finding appropriate Lagrange multipliers ${\bf z}$, ${\bm {\mathcal Z}}$, $\Lambda$, and ${\bm\epsilon}$ by requiring $ \frac{\dd {\mathscr L}}{\dd {\bm \kappa}} = \frac{\dd {\mathscr L}}{\dd {\bm \kappa}^*}  = \frac{\dd {\mathscr L}}{\dd {\bf f}} = {\bf 0}$. Henceforth, we will only consider ${\bm\kappa}^*$ derivatives, as ${\bm\kappa}$ derivatives are just their complex conjugates.

The orbital rotation on a projected wave function is expressed simply by $\hat P e^{\hat \kappa}|\Psi\ket$ with 
\begin{align}
\hat \kappa = \sum_{pq} \kappa_{pq} \hat E_{pq},
\end{align} 
where $|\Psi\ket$ is the underlying broken-symmetry wave function, and both anti-symmetric matrix ${\bm\kappa}$ and $\hat E_{pq}=a^\dag_p a_q$ are spin-dependent.
The orbital gradient and  {\bf f} gradient  of $\calE$ are hence defined by
\begin{align}
&L_{pq} =\left. \frac{\dd \calE}{\dd \kappa_{pq}^*}\right|_{{\bm\kappa}={\bf 0}}, \\% &= \bra \Psi | \hat E_{tu} \hat H \hat P |\Psi\ket - E \bra \Psi |\hat E_{tu} \hat P |\Psi\ket.
&L_{\cal K} = \left. \frac{\dd \calE}{\dd f_{\cal K}}\right|_{{\bf f}={\bf 0}},
\end{align}
and, together with the derivatives of the second and last terms of Eq.(\ref{eq:Lag}), they will constitute the Z-vector equations for post-PHF, which we refer to as CPPHF equations.
Note that $\Lambda$ is easily identified as a constant $\calE$, which can be verified by multiplying $\frac{\dd {\mathscr L}}{\dd f_{\cal K}}$ by $f_{\cal K}$ and then by summing over all ${\cal K}$.

Once {\bf z}, ${\bm {\mathcal Z}}$, and ${\bm \epsilon}$ are all determined, the energy gradient for perturbation $x$ becomes
\begin{align}
\calE^x & = \calE^{(x)} + \sum_{ia} z_{ai}  \left(\calF_{ai}^{(x)} + \calF_{ia}^{(x)} \right) \bre
&+ \sum_{\cal KK'}\Bigl[ {\cal Z}_{\cal K} \left({\cal H}^{(x)}_{\cal KK'} - E_{\rm PHF}^{(x)} {\cal N}_{\cal KK'} - E_{\rm PHF} {\cal N}_{\cal KK'} ^{(x)}\right)f_{\cal K'}  \bre 
& - \calE  {\cal N}^{(x)}_{\cal KK'} f_{\cal K} f_{\cal K'}   \Bigr]-  \sum_{pq} \epsilon_{pq} (\CSC)_{qp},\label{eq:Ex1}
\end{align} 
where superscripts $x$ and $(x)$ respectively indicate the total derivative and partial derivative with ${\bC}[{\bm\kappa}]$ fixed. 

If $\calE$ is not variational with respect to ${\bf c}$, we also have to treat {\bf c} in a similar way to {\bf C} as above. This happens when $\calE$ has an additional correction term like the Davidson correction to CI,\cite{Langhoff74,Davidson74,Duch94} or when $\calE$ and {\bf c} are determined by a projective way as in coupled-cluster (CC). In such cases, one also needs to treat {\bf c} in a similar way to {\bf C}. While the computational cost will surely increase, its formulation is straightforward, as shown in Section \ref{sec:ECISD+Q} for ECISD+Q.

As expected, the above scheme closely resembles the one for regular post-HF methods. The main difference, however, lies in the {\it density matrices}. The post-PHF density matrices in a molecular orbital (MO) basis depend not only on {\bf c} {\it but also on $x$ explicitly through {\bf S} and implicitly through {\bf C}}. This is not the case in the traditional schemes, where density matrices in a certain MO basis are only a function of {\bf c} because a transition density matrix element $\bra \Phi_I | a_p^\dag a_q | \Phi_J\ket$ is always either 0 or 1. In Eq.(\ref{eq:Ex1}), $\calE^{(x)}$ therefore requires a special treatment, which we will discuss in Section \ref{sec:Explicit}. Also, $L_{pq}$ defined as above is of the broken-symmetry representation (thus it has $\alpha$ and $\beta$ components) and is not the same as the standard generalized Fock matrix,\cite{Helgaker00} which has been extensively used as orbital gradients.
In other words, 
\begin{align}
{\mathscr P}_{pq} L_{pq} \ne {\mathscr P}_{pq}\Bigl( \sum_{r} h_{pr} P_{rq} + \sum_{rst} \bra p t| rs \ket P_{qt,rs} \Bigr),\label{eq:genF}
\end{align}
where ${\mathscr P}_{pq} = 1 - (p \leftrightarrow q)$ is the permutation operator and {\bf P} are the unrelaxed density matrices of the method in question. 
This is essentially due to the same reason as above, that is, the unrelaxed density matrices for the correlated wave function are not solely determined by {\bf c} but they depend on the MO coefficients. This somewhat complicates our derivation as will be seen. 

One could entirely avoid the broken-symmetry picture by using the internally-contracted spin-free basis where $\exp(\hat \kappa)$ is placed {\it after} the projection operator, i.e. ${\bra \Psi | \hat P {e^{\hat \kappa}}^\dag \hat H  e^{\hat \kappa}\hat P |\Psi\ket} /  \bra \Psi | \hat P {e^{\hat \kappa}}^\dag   e^{\hat \kappa}\hat P |\Psi\ket$. This is possible provided that $\hat \kappa$ is spin-free, written in, for example, the natural orbital (NO) basis. Formulated in this way, the equality between the left hand and right hand sides of Eq.(\ref{eq:genF}) is now satisfied, if {\bf P} are also spin-free. However, the broken-symmetry representation offers numerous advantages over a spin-free basis in both the derivation and implementation of analytical gradient. First, the SCF condition for PHF is given in the broken-symmetry basis (Eq.(\ref{eq:Fai})), and hence {\bf z} is also effectively spin-polarized. Second, the nonorthogonal Wick theorem makes it easy to evaluate the required matrix elements as we have previously shown.\cite{Tsuchimochi16B} Third, there is a clear distinction between occupied and virtual blocks, which enables us to separate the working spaces and to reduce the computational effort. Fourth, there is no need to deal with the double-integration that arises due to the presence of two projection operators, whose computational cost is an order of magnitude higher. Of course, one can reformulate the spin-free formalism within the single integration by adopting the well-known Wigner-Eckart theorem\cite{Tinkham,Garza14,Uejima16}; however, all the expressions will ultimately become a linear combination of the half-projected, broken-symmetry representation, so from the algebraic point of view,  it seems pointless to employ the spin-free basis. Indeed, in most situations, we only need half-projected quantities, and therefore we could utilize the Wigner-Eckart theorem only when necessary, as will be done for the spin-adapted relaxed density matrix. Finally, it is expected that most ingredients that will be needed for analytical derivatives are readily available in the broken-symmetry basis  from the existing post-PHF programs and therefore are likely to require only minor modifications. 

In what follows, we apply the above scheme to derive the ECISD nuclear gradient. We abbreviate the collinear-spin projection operator as $\hat P \equiv \hat {\cal P}^{\cal J}_{\cal MM}$, as the spin quantum number $\cal J$ and multiplicity $\cal M$ are both obvious. Also note that there is no dependence on $f_{\cal K}$ and one can remove the corresponding terms from ${\mathscr L}$ in Eq.(\ref{eq:Lag}). Again, once the derivative is given for ECISD, similar derivations can be obtained for ECIS and TDPHF excited states as well as spin-flip ECIS. For EMP2, care must be taken because it uses UHF-like orbital energies,\cite{Tsuchimochi14} but its derivation essentially follows along the same lines.\cite{Jorgensen88} 

\subsection{Nomenclature}\label{sec:Nomen}
Before going into detailed derivations, we shall briefly summarize our nomenclature to make this paper self-contained. We basically follow the same notations with Ref.[\onlinecite{Tsuchimochi16B}], but specifically clarify them here as well. 

Throughout this work, we will employ the projection operator in the following form:
\begin{align}
\hat P = \int \hat R(\Omega) d\Omega \approx \sum_g w_g \hat R_g, \label{eq:Ps}
\end{align} 
where $\hat R_g$ and $w_g$ are a rotation operator at some discretized grid point $g$ and its weight. Hence, one can generically work on the transition elements between unrotated and rotated states such as $\bra \Theta|$ and $\hat R_g|\Theta\ket = |\Theta_g\ket$, followed by the summation over $g$.

\subsubsection{SUHF}
Using  Eq.(\ref{eq:Ps}), the SUHF energy is given by
\begin{subequations}
\begin{align}
E_{\rm SUHF} = \frac{\bra \Phi|\hat H \hat P|\Phi\ket}{\bra \Phi|\hat P |\Phi\ket} =\frac{\sum_g w_g n_g E_g} {\sum_g w_g n_g},
\end{align}
where we have used the fact that $\hat P$ is Hermitian, idempotent, and commutable with spin-free operators such as $\hat H$, and defined
\begin{align}
&E_g \equiv \frac{\bra \Phi| \hat H |\Phi_g\ket}{\bra \Phi|\Phi_g\ket} = \bra \Phi| \hat H |\Phi_g\ket_N,\label{eq:Eg0}\\
&n_g \equiv \bra \Phi|\Phi_g\ket.
\end{align} 
\end{subequations}
Hereinafter we will simplify notations by adding subscript $N$ to transition elements in order to indicate the elements are intermediate-normalized, i.e., divided by $n_g$ as in Eq.(\ref{eq:Eg0}). %We may also omit subscript $g$ when its dependence is apparent.
The SUHF transition one-particle density matrix (1PDM) $\pg$ and transition Fock matrix $\Fg$ defined by
\begin{align}
&(\pg)_{pq} =\bra \Phi|a^\dag_q a_p|\Phi_g\ket_N,\\
&(\Fg)_{pq} = h_{pq} + \sum_{rs} \bra pr||qs\ket (\pg)_{sr},
\end{align} 
are fundamental quantities for our discussion below. For example, the SUHF transition energy is expressed as
\begin{align}
E_g = \sum_{pq} h_{pq} (\pg)_{qp} + \frac{1}{2} \sum_{pqrs}\bra pr||qs\ket (\pg)_{qp} (\pg)_{sr}. \label{eq:Eg1}
\end{align}
For more details, we refer the reader to the original PHF paper (Ref.[\onlinecite{Jimenez12}]).

\subsubsection{Nonorthogonal Wick theorem}
The nonorthogonal Wick theorem\cite{Tsuchimochi16A} allows one to write Hamiltonian {\it at any grid} $g$ as
\begin{align}
 \hat H = E_g + \sum_{pq} (\Fg)_{pq} \{a_p^\dag a_q \}_g  + \frac{1}{4} \sum_{pqrs} \bra pq||rs \ket \{ a_p^\dag a_q^\dag a_s a_r\}_g, \label{eq:NormalH}
\end{align}
where curly brackets mean the normal-ordering in the sense of the nonorthogonal Wick theorem with respect to the left- and right-vacua $\bra \Phi|$ and $|\Phi_g\ket$. In other words, $\bra \Phi| \{ \hat O\}_g  |\Phi_g\ket_N \equiv 0$. Then, it proves convenient to define the following quantities in the MO basis:
\begin{subequations}
\begin{align}
%&(\Wg)_{ij} = \bra \Phi| a^\dag_j \hat R_g a_i |\Phi\ket_N = (\bC_o^\dag \Rg \bC_o)^{-1}_{ij} \\
%&(\Wg)_{ai} =  \bra \Phi| a^\dag_i a_a \hat R_g |\Phi\ket_N = (\bC_v^\dag \Rg \bC_o\Woo)_{ai}\\
%&(\Wg)_{ia} =  \bra \Phi| \hat R_g a^\dag_a a_i |\Phi\ket_N = (\Woo \bC_o^\dag \Rg \bC_v)_{ia} \\
%&(\Wg)_{ab} =  \bra \Phi| a_a \hat R_g a_b^\dag |\Phi\ket_N = (\bC_v^\dag \Rg \bC_v)_{ab} -  (\bC_v^\dag \Rg \bC_o \Woo \bC_o^\dag \Rg \bC_v)_{ab}
&(\Wg)_{ij} = \bra \Phi| a^\dag_j \hat R_g a_i |\Phi\ket_N = (\Roo)^{-1}_{ij}, \\
&(\Wg)_{ai} =  \bra \Phi| a^\dag_i a_a \hat R_g |\Phi\ket_N = \left(\Rvo(\Roo)^{-1}\right)_{ai},\\
&(\Wg)_{ia} =  \bra \Phi| \hat R_g a^\dag_a a_i |\Phi\ket_N = \left((\Roo)^{-1}\Rov\right)_{ia}, \\
&(\Wg)_{ab} =  \bra \Phi| a_a \hat R_g a_b^\dag |\Phi\ket_N = \left(\Rvv - \Rvo(\Roo)^{-1}\Rov\right)_{ab},
\end{align}
\label{eq:calW}
\end{subequations}
where ${\rm o}$ and ${\rm v}$ \black stand for the occupied and virtual orbital blocks, respectively. The above equations manifest
\begin{align}
\pg = \begin{pmatrix}
{\bf 1}^{\rm oo} & {\bf 0}^{\rm ov}\\ \Wvo & {\bf 0}^{\rm vv}
\end{pmatrix}.\label{eq:rho_g}
\end{align} 
In the nonorthogonal Wick theorem, Eqs.(\ref{eq:calW}-\ref{eq:rho_g}) are realized as the contractions of two different Fermion operators, $a^\dag$ and $\hat R_g a^\dag \hat R_g^{-1}$, between $\bra \Phi|$ and $|\Phi_g\ket$. Note that all other possible contractions result in either the Kronecker delta or zero. Then, the standard Wick theorem and its generalization for a product of normal-ordered strings can be completely replaced by their generalization to nonorthogonal bases. 
 
In the previous work,\cite{Tsuchimochi15C,Tsuchimochi16A,Tsuchimochi16B} we introduced the left- and right-transformation matrices, given by
\begin{align}
\Lg  = \begin{pmatrix}
\Woo & {\bf 0}^{\rm ov}\\ -\Wvo & {\bf 1}^{\rm vv}
\end{pmatrix}, \\
\R  = \begin{pmatrix}
{\bf 1}^{\rm oo} & {\bf 0}^{\rm ov}\\ \Wvo & \Wvv
\end{pmatrix},
\end{align}
to manipulate the transition Fock matrix and the bare two-electron integrals as
\begin{align}
&\tilde {\bm \calF}_{g} := \Lg \Fg \R,\\
&(\calV_g)_{rs}^{pq} := \sum_{tuvw} (\Lg)_{pt}(\Lg)_{qu} \bra tu||vw\ket (\R)_{vr} (\R)_{ws}. \label{eq:Vg}
\end{align} 
It is sometimes convenient to treat these matrices together with the MO transformation in some cases, 
\begin{align}
\tilde {\bm\calL}_g = \Lg \bC^\dag,\\
\tilde {\bm\calR}_g = \bC \R,
\end{align}
so that both $\calE$ and $\calF_{ai}$ are solely expressed as a function of $\Wg$, $\tilde {\bm\calL}_g$, and $\tilde {\bm\calR}_g$, without explicit ${\bf C}[{\bm\kappa}]$ dependence. 

\subsubsection{ECISD energy and density matrices}
Similarly to the SUHF energy, the ECISD energy can be expressed as
\begin{subequations}
\begin{align}
\calE =  \frac{ \sum_g w_g n_g \calE_g} {\sum_g w_g n_g \calN_g}, \label{eq:EECISD1}
\end{align} 
where we have defined the transition energy and overlap
\begin{align}
\calE_g& = \bra \Psi| \hat H | \Psi_g\ket_N \bre
&= \sum_{\mu\nu} h_{\mu\nu} \bra \Psi | a_\mu^\dag a_\nu |\Psi_g\ket_N + \frac{1}{4} \sum_{\mu\nu\lambda\sigma}\bra \mu\nu||\lambda\sigma\ket  \bra \Psi | a_\mu^\dag a_\nu^\dag a_\sigma a_\lambda |\Psi_g\ket_N,\label{eq:calEg}\\
\calN_g& = \bra \Psi | \Psi_g\ket_N,\label{eq:calNg}
\end{align}
\label{eq:EECISDx}
\end{subequations}
for which, we have presented the expression of $\sigma$-vectors in the previous work. We will not repeat their derivations but the supporting material is available for the final expressions, which are factorized and thus more compact than those presented in Refs.[\onlinecite{Tsuchimochi16A,Tsuchimochi16B}].

In what follows, we assume both $\bra \Psi|\hat P |\Psi\ket$ and $\bra \Phi |\hat P|\Phi\ket$ are properly normalized, i.e., $\sum_g w_g n_g \calN_g = \sum_g w_g n_g = 1$, for brevity; however, we stress that these norms need be taken into account in derivative evaluations (as there derivatives are typically nonzero).

Note that half-projected density matrices, such as $ P_{\nu\mu} \equiv \bra \Psi | a_\mu^\dag a_\nu \hat P |\Psi\ket = \sum_g w_g n_g \bra \Psi | a_\mu^\dag a_\nu | \Psi_g\ket_N$, are neither relaxed nor spin-adapted. As these ``unrelaxed'' ECISD density matrices will repeatedly appear in the following derivations, it is useful to analyze these objects for latter discussions. 1PDM $P_{qp}$ is obtained simply by integrating the corresponding transition 1PDM, ${\bf P}_g$,
\begin{align}
P_{qp} = \sum_g w_g n_g ({\bf P}_g)_{qp} = \sum_g w_g n_g \bra \Psi | a_p^\dag a_q |\Psi_g\ket_N,
\end{align} 
and similarly for two-particle density matrix (2PDM). 
The nonorthogonal Wick theorem applied to the one-particle operator
\begin{align}
a_p^\dag a_q = (\pg)_{qp} +\{ a_p^\dag  a_q \}_g  \label{eq:Epq}
\end{align} 
suggests ${\bf P}_g$ be separated into two terms, namely, the overlap-weighted SUHF contribution $(\pg)_{qp} \calN_g$ and the normal-ordered transition 1PDM defined by
\begin{align}
({\bm\gamma}_g)_{pq} = \bra \Psi| \{a_q^\dag a_p\}_g |\Psi_g \ket_N,\label{eq:gamma}
\end{align} 
which accounts for the correction due to the ECISD correlation contribution. The programmable expression of Eq.(\ref{eq:gamma}) is easily identified from the ECISD equations, as ${\bm\gamma}_g$ is contracted only with $\Fg$ because of the structure of the normal-ordered Hamiltonian Eq.(\ref{eq:NormalH}). Namely, it suffices to replace all $(\tilde{\bm\calF}_g)_{rs}$ with $(\R)_{rp}(\Lg)_{qs}$ and neglect all other contractions. Note that the correlated (non-separable) 2PDM contribution, $\bra \Psi|\{a_p^\dag a_q^\dag a_s a_r\}_g |\Psi_g\ket_N$, is similarly obtained, but with $({\bm\calV}_g)_{rs}^{pq}$ replaced appropriately (see Eq.(\ref{eq:Vg})).

\subsection{ECISD orbital gradient}\label{sec:ECISDFock}
Having established our notations above, now we are in a position to derive the ECISD orbital gradient {\bf L}. Here the generalized nonorthogonal Wick theorem proves useful. Using
Eq.(\ref{eq:Epq})
and the normal ordered Hamiltonian Eq.(\ref{eq:NormalH}), we write
\begin{align}
L_{pq}% &= \frac{\dd E}{\dd \kappa_{ut}^*}  \bre 
&= \bra \Psi|\hat E_{pq}^\dag (\hat H -\calE) \hat P| \Psi\ket \bre
& = \sum_g w_g n_g \Biggl[  (\pg)_{pq} \bra \Psi | \hat H |\Psi_g\ket_N + \bra \Psi | \{a_q^\dag a_p\}_g | \Psi_g\ket_N E_g  \bre
&+ \bra \Psi | \{a_q^\dag a_p\}_g\{a_r^\dag a_s\}_g | \Psi_g\ket_N (\Fg)_{rs} \bre
&+ \frac{1}{4} \bra rs||tu\ket \bra \Psi | \{a_q^\dag a_p\}_g \{ a^\dag_r a^\dag_s a_u a_t\}_g | \Psi_g\ket_N - ({\bf P}_g)_{pq} \calE\Biggr] \bre
 &= \sum_g w_g n_g \Biggl[({\bf P}_g)_{pq} (E_g  - \calE) +(\pg)_{pq} (\calE_g - E_g \calN_g) \bre
 &- [{\bm\gamma}_g\Fg{\bm\rho}_g]_{pq} +  [{\bm\eta}_g\Fg{\bf P}_g]_{pq}+  [{\bm\eta}_g{\bm {\mathcal G}}_g \pg]_{pq} \bre
 &+ ({\bm \zeta}_g)_{pq} -  ({\bm \omega}_g)_{pi}\delta_{iq} + \delta_{pa}(\tilde {\bm \omega}_g)_{aq} \Biggr]. \label{eq:ECIFock}
\end{align} 
Hereafter, the Einstein summation convention for repeated indices is assumed for the sake of visual simplification, except $g$ integration in order to specifically indicate that the symmetry-projection is being performed. In Eq.(\ref{eq:ECIFock}), we have additionally defined the transition hole matrix
\begin{align}
({\bm\eta}_g)_{pq} = \delta_{pq} - (\pg)_{pq},
\end{align} 
and the following $g$-dependent quantities:
%%%%%%%%
%      Gg        %
%%%%%%%%
\begin{align}
({\bm{\mathcal G}}_g )_{pq} :&= \bra pr||qs\ket ({\bm \gamma}_g)_{sr}, \label{eq:Gg}\\
({\bm \zeta}_g)_{pq} :&= (\Fg)_{rs}\bra \Psi | \{a_q^\dag a_r^\dag a_s a_p \}_g | \Psi_g\ket_N  \bre
%&+ \frac{1}{4}\bra pq||rs\ket \bra \Psi | \{a_t^\dag a_p^\dag a_q^\dag a_s a_r a_u \}_g | \Psi_g\ket_N \label{eq:zeta}\\
&+ \frac{1}{4}\bra rs||tu\ket \bra \Psi | \{a_q^\dag a_r^\dag a_s^\dag a_u a_t a_p \}_g | \Psi_g\ket_N,\label{eq:zeta}\\
%({\bm \omega}_ g)_{ut}:&= \frac{1}{2}\bra pq||rs\ket (\R)_{rt}\bra \Psi | \{  a_p^\dag a_q^\dag a_s a_u \}_g | \Psi_g\ket_N\label{eq:omega}\\
({\bm \omega}_ g)_{pq}:&= \frac{1}{2}\bra rs||tu\ket (\R)_{tq}\bra \Psi | \{  a_r^\dag a_s^\dag a_u a_p \}_g | \Psi_g\ket_N,\label{eq:omega}\\
%(\tilde {\bm\omega}_g)_{ut} :&= \frac{1}{2}\bra pq||rs\ket  (\Lg)_{up}\bra \Psi | \{a_t^\dag a_q^\dag a_s a_r\}_g | \Psi_g\ket_N.\label{eq:tomega} 
   (\tilde {\bm\omega}_g)_{pq} :&= \frac{1}{2}\bra rs||tu\ket  (\Lg)_{pr}\bra \Psi | \{a_q^\dag a_s^\dag a_u a_t\}_g | \Psi_g\ket_N.\label{eq:tomega} 
\end{align}
This is a general result for any excitation levels, including SUHF and ECIS. For example, {\bf L} will reduce to ${\bm{\calF}}$ in the case of SUHF. For ECISD, the explicit expressions of Eqs.(\ref{eq:zeta}-\ref{eq:tomega}) are rather complex as given in the supplemental material, but can be straightforwardly evaluated using the existing ECISD subroutines with minor modifications. We just note here that we avoid the computation of the correlated 3PDM in Eq.(\ref{eq:zeta}) by performing the integral contraction on-the-fly, and ${\bf L}$ can be thus evaluated for the same cost as the ECISD energy, which scales as ${\cal O}({\rm o}^2 {\rm v}^4)$. 

Some observations on the structure of {\bf L} are in order. Since the ECISD energy is invariant with respect to an orbital rotation within the occupied space as well as within the virtual space, clearly $L_{ij} = L_{ab} = 0$. Furthermore, $L_{ia}$ must also vanish because it apparently constitutes only $\bra \Phi | (\hat H - E) \hat P |\Psi\ket$ and $\bra \Phi_i^a | (\hat H - E) \hat P |\Psi\ket$, which are all guaranteed to be zero due to the variationality of ECISD with respect to the CI coefficients. Therefore, only the $L_{ai}$ block contains nonzero elements, as confirmed numerically in our calculations.

\subsection{Coupled-perturbed PHF}\label{sec:CPPHF}
With {\bf L} derived above, we also need the orbital derivative of SUHF Fock matrix (\ref{eq:Fai}) for computing $z_{ai}$. Using the notations defined in Section \ref{sec:Nomen}, $\calF_{ai}$ is explicitly written as
\begin{align}
\calF_{ai} = \frac{\dd E_{\rm SUHF}}{\dd \kappa_{ai}^*} = \sum_{g}w_g n_g \Bigl[\left(E_g - E_{\rm PHF}\right)\Wg + \Lg\Fg\Rg \Bigr]_{ai},
\end{align}
and ${\calF}_{ia}$ is its complex conjugate.
As previously shown,\cite{Tsuchimochi15A} its derivative with respect to an orbital rotation becomes the Hessian components,
%%%%%%%%%%%
%         Fai,  kj         %
%%%%%%%%%%%
\begin{subequations}
\begin{align}
 \frac{\dd \calF_{ia}}{\dd \kappa_{bj}^*} &= A_{ai,bj}\bre
 % &= \sum_g w_g n_g \Bigl[(E_g - E_{\rm PHF}) \Bigl( (\Wg)_{ai}  (\Wg)_{jb} +  (\Wg)_{ab} (\Wg)_{ji} \Bigr)+  (\Wg)_{ai}  (\btcalF)_{jb} + (\btcalF)_{ai}  (\Wg)_{jb} \bre
%& + (\btcalF)_{ab}  (\Wg)_{ji} -  (\Wg)_{ab} (\btcalF)_{ji} + (\calV_g)_{ib}^{aj} \Bigr]\\
% \frac{\dd \calF_{ai}}{\dd \kappa_{bj}^*} = B_{ai,bj} &= \sum_g w_g n_g \PP(ab)\PP(ij) \Bigl[  \frac{1}{2} (E_g - E_{\rm PHF})   (\Wg)_{ai} (\Wg)_{bj} +  (\Wg)_{ai}(\btcalF)_{bj} + \frac{1}{4} (\calV_g)^{ab}_{ij}\Bigr]
 &= \sum_g w_g n_g \Bigl[(E_g - E_{\rm SUHF}) \Bigl( \calW_{ai}  \calW_{jb} +  \calW_{ab} \calW_{ji} \Bigr) \bre
 &+  \calW_{ai}  \tcalF_{jb} + \tcalF_{ai}  \calW_{jb}  + \tcalF_{ab}  \calW_{ji} -  \calW_{ab} \tcalF_{ji} + \calV_{ib}^{aj} \Bigr],\\
 \frac{\dd \calF_{ai}}{\dd \kappa_{bj}^*} &= B_{ai,bj}\bre
  &= \sum_g w_g n_g \PP(ab)\PP(ij) \Bigl[  \frac{1}{2} (E_g - E_{\rm SUHF})   \calW_{ai} \calW_{bj} \bre
  &+  \calW_{ai}\tcalF_{bj} + \frac{1}{4} \calV^{ab}_{ij}\Bigr],
\end{align}
\label{eq:AB}
\end{subequations}
where we have dropped $g$ subscripts in $\Wg$, $\tilde{\bm  \calF}_g$, and $\bar {\bm{\mathcal V}}_g$ for brevity. All the derivatives with respect to $\kappa_{pq}^*$  in the ${\rm oo}$, ${\rm ov}$, and ${\rm vv}$ spaces are rigorously zero for a similar reason to the aforementioned discussion for {\bf L}. In deriving Eqs.(\ref{eq:AB}), we have used the fact that $ {\bm\calF} = {\bf 0}$  for a converged SUHF state.

Hence, the stationary condition $\frac{\dd {\mathscr L}}{\dd {\bm\kappa}^*} = {\bf 0}$ reads the following set of equations;
\begin{subequations}
\begin{align}
& L_{ij} = \epsilon_{ij},\\
& L_{bj} +  \left( A_{ai,bj} + B_{ai,bj} \right) z_{ai} = \epsilon_{bj},\label{eq:CPPHF0}\\
& L_{jb}= \epsilon_{jb},\\
& L_{ab} = \epsilon_{ab}.
\end{align}  
\label{eq:epq}
\end{subequations}
Keeping the structure of {\bf L}  in mind (${\bf L}^{\rm vo}$ is the only nonzero block)  and requiring ${\bm\epsilon} = {\bm \epsilon}^\dag$, we find simply ${\bm\epsilon} = {\bf 0}$ as is the case in SUHF.\cite{Schutski14} 
On the other hand, Eq.(\ref{eq:CPPHF0}) results in the Z-vector (CPPHF) equation,
%\begin{itembox}{{\red Coupled-perturbed PHF equation}}
\begin{align}
\left(A_{ai,bj}+B_{ai,bj}\right) z_{ai}=  - L_{bj}.\label{eq:CPPHF}
\end{align} 
%\end{itembox} 
Again, for general projection operators in the form of Eq.(\ref{eq:generalP}), this equation will be coupled with the corresponding {\bf f} response, similarly to MRCI.\cite{Rice85}

Orbital rotations can often contain linear dependencies, i.e., Hessian ${\bf A} + {\bf B}$ is singular in PHF, since a projection operator may produce the identical symmetry-adapted state from different broken-symmetry determinants.\cite{Tsuchimochi15A} If this is the case, such linear dependencies also appear in {\bf L} in exactly the same way, so one can easily identify this redundant space as a mathematically {\it null} space. 
The reader may then wonder if {\bf z} is left arbitrary and there are infinite numbers of solutions that satisfy the Z-vector equation (\ref{eq:CPPHF}).
However, this redundancy is simply due to the working space that we have adopted; had we chosen an appropriate space for orbital rotations other than the broken-symmetry representation, such a null space would completely disappear from the equation. Therefore, the correct approach to treat these linear dependencies in the present Z-vector equation is to simply remove them; in practice, we take the pseudo-inverse of ${\bf A}+{\bf B}$ with a threshold of $10^{-9}$ in order to ensure a numerical stability. This procedure determines only one unique {\bf z}. In passing, if other {\bf z} that yet satisfy Eq.(\ref{eq:CPPHF}) are used, gradients would still be correctly evaluated but the relaxed density matrix may become different.

For ECISD, one can solve Eq.(\ref{eq:CPPHF}) by explicitly forming and pseudo-inverting the Hessian matrix ${\bf A} + {\bf B}$, because its cost is ${\cal O}({\rm o}^3 {\rm v}^3)$, which is typically less than that of ECISD itself. For other low-scaling methods such as EMP2, one should resort to iterative linear-equation solvers like GMRES for efficient computations.

\ifx10
\subsection{Frozen-core approximation}
In ECISD, the frozen-core approximation becomes a little tricky, as it employs a constrained optimization in PHF. This is very similar to that of constrained unrestricted HF of Tsuchimochi and Scuseria, where the Fock matrix is modified. In this scheme, we divide the PHF orbital space into the core ($C$), active ($A$), and virtual spaces ($V$). Note that $V$ is different from $v$, and is intrinsic virtual orbitals  whose occupation is zero. Here only the orbitals in the active space are symmetry-broken, while $C$ and $V$ spaces are forced to preserve the symmetry and require some constraint. The constrained Fock is then written as
\begin{align}
\tilde \calF_{ia}^\alpha = \tilde \calF_{ia}^{\beta} = \frac{1}{2} \left( \calF_{ia}^\alpha+  \calF_{ia}^\beta\right) \qquad i\in C, a\in V
\end{align} 
 Hence, one needs a slight modification to the PHF-SCF condition for the $CV$ and $VC$ spaces. The constraint for Lagrangian is actually
\begin{align}
& \sum_{ai}^{AC} z^{AC}_{ai} \left(\calF^\alpha_{ia} + \calF^\alpha_{ai}\right) + \sum_{ai}^{AC} \bar z^{AC}_{ a i} \left(\calF^\beta_{ ia} + \calF^\beta_{ a i}\right) \bre
&+ \sum_{ai}^{VA} z^{VA}_{ai}\left(\calF^\alpha_{ia} + \calF^\alpha_{ai}\right) + \sum_{ai}^{VA} \bar z^{VA}_{ a i} \left(\calF^\beta_{ia} + \calF^\beta_{ a i}\right) \bre 
&+\frac{1}{2} \sum_{ai}^{VC}  \left(z^{VC}_{ai} + \bar z_{ai}^{VC}\right) \left(\calF^\alpha_{ia} + \calF^\beta_{ i a} + \calF^\alpha_{ai} + \calF^\beta_{ ai}\right) \bre 
\end{align}
The orbital gradient of the constrained Fock is

\fi

\subsection{Explicit dependence}\label{sec:Explicit}
\subsubsection{$\calE^{(x)}$}
In Section \ref{sec:General}, we argued that the explicit dependence on $x$ in $\calE$ comes not only from Hamiltonian but also from density matrices through {\bf S}. Given the ECISD energy Eqs.(\ref{eq:EECISDx}), one would have to consider terms like $\bra \Psi| a_p^\dag a_q | \Psi_g\ket_N^{(x)}$ and  $\bra \Psi| a_p^\dag a^\dag_q a_s a_r | \Psi_g\ket_N^{(x)}$, which are both exactly zero if $|\Psi_g\ket = |\Psi\ket$ (i.e., the regular single-reference limit) but are complicated functions of both ${\bf C}$ and ${\bf S}^x$ in the symmetry-projection methods. However, one can completely avoid constructing these derivatives and simplify the derivation by formally writing the explicit dependence of $\calE$ on $x$ as
\begin{align}
\calE^{(x)} &=  \frac{\dd  \calE}{\dd h_{\mu\nu}} h^x_{\mu\nu} +  \frac{\dd  \calE}{\dd \bra \mu\nu||\lambda\sigma\ket }\bra \mu\nu||\lambda\sigma\ket^x \bre
&+ \sum_g \Bigl(  \frac{\dd  \calE}{\dd  {\bm \calW}_g} {\bm \calW}_g^{(x)}+ \frac{\dd  \calE}{\dd \tilde {\bm \calL}_g}  \tilde{\bm \calL}_g^{(x)}+ \frac{\dd  \calE}{\dd \tilde{\bm \calR}_g}  \tilde{\bm \calR}_g^{(x)}\Bigr).\label{eq:E(x)}
\end{align}
Then, the last summation over $g$ takes into account the fact that the density matrices of ECISD are dependent on $x$. 
Now, recall that we require $\calE$ be completely expressed with  ${\bm \calW}_g$, $ \tilde {\bm \calL}_g$, and $ \tilde {\bm \calR}_g$, and hence allow no explicit dependence on ${\bf C}$. Therefore, the previously obtained ECI orbital gradient Eq.(\ref{eq:ECIFock}) is equivalently expressed as
\begin{align}
L_{pq} = \frac{\dd \calE}{\dd \kappa_{pq}^*}=   \sum_g \Bigl( \frac{\dd  \calE}{\dd  {\bm \calW}_g} \frac {\dd {\bm \calW}_g}{\dd \kappa_{pq}^*}+ \frac{\dd \calE}{\dd \tilde {\bm \calL}_g}  \frac{\dd\tilde{\bm \calL}_g}{\dd \kappa_{pq}^*}+ \frac{\dd \calE}{\dd \tilde{\bm \calR}_g}  \frac{\dd \tilde{\bm \calR}_g}{\dd \kappa_{pq}^*}\Bigr),\label{eq:Fock1}
\end{align} 
which may be compared with Eq.(\ref{eq:E(x)}) for their similarity, implying that the latter can be rewritten in terms of $L_{pq}$.
It is not difficult to show that the explicit derivatives of ${\bm \calW}_g$, $ \tilde {\bm \calL}_g$, and $ \tilde {\bm \calR}_g$ with respect to $x$ critically resemble those with respect to $\kappa_{pq}^*$,
%\begin{screen}
\begin{subequations}
\begin{align}
 {\bm\calW}_g^{(x)} &= \frac{\dd {\bm\calW}_g}{\dd \kappa_{pq}^*}(\CSC)_{qp},\\
\tilde {\bm\calL}_g^{(x)}& = \frac{\dd \tilde{\bm \calL}_g}{\dd \kappa_{pq}^*}(\CSC)_{qp} - \tilde{\bm\calL}_g\bS^x \bS^{-1},\\
 \tilde {\bm\calR}_g^{(x)} &= \frac{\dd \tilde{\bm \calR}_g}{\dd \kappa_{pq}^*}(\CSC)_{qp}.
\end{align}
\label{eq:x_and_kappa}
\end{subequations}
%\end{screen}
Therefore, noting that only $\calE_g$ depends on $\tilde {\bm \calL}_g$ in $\calE$, one can substitute them into Eq.(\ref{eq:E(x)}) and then use Eqs.(\ref{eq:epq},\ref{eq:Fock1}) to obtain
\begin{align}
\calE^{(x)} 
%&=  \frac{\dd  E}{\dd h_{\mu\nu}} h^x_{\mu\nu} +  \frac{\dd  E}{\dd \bra \mu\nu||\lambda\sigma\ket }\bra \mu\nu||\lambda\sigma\ket^x +    \sum_g \Biggl( \frac{\dd  E}{\dd  {\bm \calW}_g} {\bm \calW}_g^{(x)}+ \frac{\dd  E}{\dd \tilde {\bm \calL}_g}  \tilde{\bm \calL}_g^{(x)}+ \frac{\dd  E}{\dd \tilde{\bm \calR}_g}  \tilde{\bm \calR}_g^{(x)}\Bigr).
%\bre
%&=\frac{\dd  E}{\dd h_{\mu\nu}} h^x_{\mu\nu} +  \frac{\dd  E}{\dd \bra \mu\nu||\lambda\sigma\ket }\bra \mu\nu||\lambda\sigma\ket^x + \sum_g \Biggl(  \frac{\dd E}{\dd {\bm \calW}_g}\frac{\dd {\bm \calW}_g}{\dd \kappa_{pq}^*} (\CSC)_{qp} + \frac{\dd E}{\dd \tilde {\bm \calL}_g} \Bigl(\frac{\dd \tilde{\bm \calL}_g}{\dd \kappa_{pq}^*} (\CSC)_{qp} - \tilde{\bm\calL}_g\bS^x \bS^{-1}  \Bigr) \bre
%&+ \frac{\dd E}{\dd \tilde{\bm \calR}_g}\frac{\dd \tilde{\bm \calR}_g}{\dd \kappa_{pq}^*}(\CSC)_{qp} \Biggr]\bre
&=  \frac{\dd  \calE}{\dd h_{\mu\nu}} h^x_{\mu\nu} +  \frac{\dd  \calE}{\dd \bra \mu\nu||\lambda\sigma\ket }\bra \mu\nu||\lambda\sigma\ket^x + L_{pq} (\CSC)_{qp} \bre
&-  \sum_g \Biggl( \frac{\dd \calE}{\dd (\tilde{\bm\calL}_g)_{r\mu}} (\tilde{\bm\calL}_g\bS^x \bS^{-1}) _{r \mu}\Biggr)\bre
%&=  L_{pq} (\CSC)_{qp} + \sum_g \Biggl( \frac{\dd  \calE_g}{\dd h_{\mu\nu}} h^x_{\mu\nu} +  \frac{\dd  \calE_g}{\dd \bra \mu\nu||\lambda\sigma\ket }\bra \mu\nu||\lambda\sigma\ket^x -\frac{\dd \calE_g}{\dd \tilde\calL_{r\mu}} ({\bm\calL}_g\bC^\dag \bS^x \bC \bC^\dag) _{r \mu}\Biggr)\bre
%&=  L_{pq} (\CSC)_{qp} + \sum_g \Biggl( \frac{\dd  \calE_g}{\dd h_{\mu\nu}} h^x_{\mu\nu} +  \frac{\dd  \calE_g}{\dd \bra \mu\nu||\lambda\sigma\ket }\bra \mu\nu||\lambda\sigma\ket^x -  \Bigl(C_{\mu p}^* \frac{\dd \calE_g}{\dd \tilde\calL_{r\mu}}\calL_{rq}\Bigr)_{pq} (\bC^\dag \bS^x \bC)_{qp} \Biggr)\bre
%&=  h_{\mu\nu}^{x} \bra \Psi | a_\mu^\dag a_\nu \hat P |\Psi\ket + \frac{1}{4} \bra \mu\nu||\lambda\sigma\ket ^{x} \bra \Psi | a_\mu^\dag a_\nu^\dag a_\sigma a_\lambda \hat P |\Psi\ket   +L_{pq} (\CSC)_{qp} - \sum_g w_g n_g \Biggl(  \Bigl(C_{\mu p}^* \frac{\dd H_g}{\dd \tilde\calL_{r\mu}}\calL_{rq}\Bigr)_{pq} (\bC^\dag \bS^x \bC)_{qp} \Biggr)\bre
&=  h_{\mu\nu}^{x} \bra \Psi | a_\mu^\dag a_\nu \hat P |\Psi\ket + \frac{1}{4} \bra \mu\nu||\lambda\sigma\ket ^{x} \bra \Psi | a_\mu^\dag a_\nu^\dag a_\sigma a_\lambda \hat P |\Psi\ket  \bre
&- (A_{ai,bj}+B_{ai,bj}) z_{ai} (\CSC)_{jb} -(\bC^\dag \bS^x \bC)_{qp} X_{pq}.\label{eq:E(x)1}
\end{align} 
The third term will cancel out the same term in ${{\bm\calF}}^{(x)}$ ({\it vide infra}).
For the last term of Eq.(\ref{eq:E(x)1}), it is relatively easy to derive
\begin{align}
X_{pq} &= \sum_g w_g n_g C_{\mu p}^* \frac{\dd \calE_g}{\dd (\tilde{\bm\calL}_g)_{r\mu}}(\Lg)_{rq}\bre
&=\sum_g w_g n_g \Bigl( [\Fg{\bf P}_g]_{pq}+ [{\bm {\mathcal G}}_g{\bm \rho}_g]_{pq}  +  [\Lg^{-1} \tilde {\bm \omega}_g]_{pq}\Bigr), \label{eq:lastterm}
\end{align}
and this will be in part recognized as the energy-weighted density matrix upon a grid integration. We should mention that the existence of $\Lg^{-1}$ is always guaranteed.
%We also note that $[\Lg^{-1} \tilde {\bm \omega}_g]_{pq}  = \frac{1}{2}\bra pt||rs\ket \bra \Psi | \{a_q^\dag a_t^\dag a_s a_r\}_g | \Psi_g\ket_N$.
%\begin{align}
%  C_{\mu p}^* \frac{\dd H_g}{\dd \tilde\calL_{r\mu}}\calL_{rq} = F_{pr} \bra \Psi| a_q^\dag a_r |\Psi_g\ket_N + \bra pr||qs\ket \bra \Psi| \{a_r^\dag a_s\}_g |\Psi_g\ket + \frac{1}{2}\bra pt||rs\ket \bra \Psi | \%{a_q^\dag a_t^\dag a_s a_r\}_g | \Psi_g\ket_N
%\end{align} 

\subsubsection{${\boldmath\calF}^{(x)}$}
One can use the same simplification as presented above to ease the evaluation of the explicit nuclear gradient contribution of the PHF Fock; ${\bm\calF}^{(x)}$ is closely related to the orbital gradients {\bf A} and {\bf B} through Eqs.(\ref{eq:x_and_kappa}). Our result is
\begin{subequations}
 \begin{align}
\calF_{ia}^{(x)}&  = \sum_g w_g n_g \Bigl[ {\bm\calL}_g \Fg^{(\bar x)} {\bm\calR}_g
  +  E_g^{(\bar x)}(\Wg- {\bf N})  \Bigr]_{ia}\bre
  &+ (\CSC)_{qp} (A_{ai,bj}\delta_{pb}\delta_{jq}  -  Y_{ia,pq}),   \\
\calF_{ai}^{(x)}&  = \sum_g w_g n_g \Bigl[ {\bm\calL}_g \Fg^{(\bar x)} {\bm\calR}_g
  +  E_g^{(\bar x)}(\Wg- {\bf N})  \Bigr]_{ai}\bre & +(\CSC)_{qp} (B_{ai,bj}\delta_{pb}\delta_{jq} - Y_{ai,pq} ), 
 \end{align} 
 \label{eq:Fx}
 \end{subequations}
 where {\bf Y} is the residual effect,
  \begin{align}
Y_{vw,pq} &= \sum_g w_g n_g C_{\mu p}^*\frac{\dd \calF_{vw}}{\dd (\tilde {\bm{\calL}}_g)_{r\mu}} (\Lg)_{rq} \bre
&= \sum_g w_g n_g \Biggl\{  [\Fg \pg]_{pq} (\Wg- {\bf N})_{vw}  \bre
&+ (\Lg)_{vq}[\Fg\R]_{pw}  + \bra pr || us\ket    (\pg)_{uq} (\Lg)_{vr} (\R)_{s w}\Biggr\},
\end{align}
and the bars on $x$ indicate that only Hamiltonian is subject to the differentiation, that is to say, in the AO basis,
 \begin{align}
&  (\Fg^{(\bar x)})_{\lambda\sigma} = h^{x}_{\lambda\sigma} + \bra \lambda\mu||\sigma\nu\ket ^{x}  (\pg)_{\nu\mu},\\
 &E^{(\bar x)}_g = h^{x}_{\lambda\sigma}   (\pg)_{\sigma\lambda}+ \frac{1}{2} \bra \lambda\mu||\sigma\nu\ket ^{x}  (\pg)_{\nu\mu}  (\pg)_{\sigma\lambda}.
 \end{align}
 (Note that $\pg$ is an explicit function of {\bf S}.)
 Notice that in Eqs.(\ref{eq:Fx}), $\Fg^{(\bar x)}$ is given in the MO basis.
 We have also introduced the SUHF norm derivatives, 
 \begin{align}
 N_{ai} & = \sum_g w_g n_g \calW_{ai} = \bra \Phi_i^a |  \hat P |\Phi\ket,\\
 N_{ia} &= \sum_g w_g n_g \calW_{ia} = \bra \Phi | \hat P  |\Phi_i^a\ket.
 \end{align} 
 As was mentioned above, ${\bf A}$ and {\bf B} terms in Eq.(\ref{eq:Fx}) are canceled out with those in $E^{(x)}$, when contracted with  $z_{ai}$. 
 
 %%%%%%
 % TABLE %
 %%%%%%
  
 \subsection{Final assembly}
 Putting altogether, we finally arrive at the complete expression,
\begin{align}
{\mathscr L}^{(x)} 
& =  \calE^{(\bar x)} + \sum_g w_g n_g \Biggl[ \left((\Lg\Fg^{(\bar x)}\R)_{ia} + (\Lg\Fg^{(\bar x)}\R)_{ai}\right)  z_{ai}
  \bre
  &
 +  E_g^{(\bar x)}\Bigl((\Wg)_{ia} -N_{ia} + (\Wg)_{ai}-N_{ai} \Bigr) z_{ai}  \Biggr]
\bre
& -  (\CSC)_{qp} \Biggl\{ X_{pq} + (Y_{ia,pq} + Y_{ai,pq}) z_{ai}  \Biggr\} \bre
&= h_{\mu\nu}^{x} P^{\rm rel}_{\nu\mu} +\frac{1}{4}  \bra \mu\nu||\lambda\sigma\ket ^{x} P^{\rm rel}_{\nu\mu,\sigma\lambda} + S^x_{\mu\nu} W_{\nu\mu}\label{eq:Lx}
%&= h_{\mu\nu}^{x} P^{\rm rel}_{\nu\mu} +  \bra \mu\nu|\lambda\sigma\ket ^{x} \tilde P^{\rm rel}_{\nu\mu,\sigma\lambda} + S^x_{\mu\nu} W_{\nu\mu}\bre
\end{align}
where $P^{\rm rel}_{\nu\mu}$ and $P^{\rm rel}_{\nu\mu,\sigma\lambda}$ are {\it spin-incomplete} relaxed one- and two-particle density matrices, and hence includes not only $\alpha\alpha$ and $\beta\beta$ but also nonzero $\beta\alpha$ and $\alpha\beta$ sectors. Using Eq.(\ref{eq:lastterm}), {\bf W} is the generalized energy-weighted density matrix explicitly given by
\begin{align}
W_{pq} % &= - \sum_g w_g n_g \Biggl\{  [\Fg \pg]_{pq} (\calW_{ia} + \calW_{ai} - N_{ia}-tilde P_{ai})z_{ai}  +[\Fg\R]_{pa} z_{ai} \calL_{iq} +[\Fg\R]_{pi} z_{ia} \calL_{aq}  \bre
%& + \bra pr || us \ket    \rho_{uq} \left(\calR_{s a}z_{ai}\calL_{ir} + \calR_{s i} z_{ia} \calL_{ar} \right)  -[\Fg \pg]_{pq} \bra \Psi|\Psi_g\ket + [\Fg {\bm\gamma}_g]_{pq} + \bra pr||us\ket \rho_{uq} \gamma_{sr} +  [\Lg^{-1} \tilde {\bm \omega}_g]_{pq}  \Biggr\}\bre
%&= - \sum_g w_g n_g \Biggl\{  [\Fg \pg]_{pq} \Bigl[ (\calW_{ia} + \calW_{ai} - N_{ia}-N_{ai})z_{ai} +  \bra \Psi|\Psi_g\ket \Bigr]  +[\Fg\R]_{pa} z_{ai} \calL_{iq} +[\Fg\R]_{pi} z_{ia} \calL_{aq}  \bre
%& + \bra pr || us \ket    \rho_{uq} \left(\calR_{s a}z_{ai}\calL_{ir} + \calR_{s i} z_{ia} \calL_{ar}\right) + [{\bm{\mathcal G}}_g \pg]_{pq} + [\Fg {\bm\gamma}_g]_{pq} +  [\Lg^{-1} \tilde {\bm \omega}_g]_{pq}  \Biggr\} \label{eq:W}
%\\
&= -\sum_g w_g n_g \Biggl[ (\Fg)_{pr} ({\bf P}_g+{\bf P}_g^{\rm corr} )_{rq}  \bre
&+ \bra pr || us \ket    (\pg)_{uq} \Bigl((\R)_{s a}z_{ai}(\Lg)_{ir} + (\R)_{s i} z_{ia} (\Lg)_{ar}\Bigr)\bre
& +[{\bm{\mathcal G}}_g \pg]_{pq} +  [\Lg^{-1} \tilde {\bm \omega}_g]_{pq}   \Biggr]
\end{align}
in the MO basis, with the relaxation correction at grid $g$ defined as
\begin{align}
%&({\bf P}_g)_{pq} = \bra \Psi | a_p^\dag a_q |\Psi_g\ket_N = (\pg)_{pg}\bra \Psi|\Psi_g\ket + (\bm\gamma)_{pq} \\
({\bf P}_g^{\rm corr})_{pq} & =  (\pg)_{pq} \Bigl((\Wg)_{ia} - N_{ia} + (\Wg)_{ai} -N_{ai}\Bigr)z_{ia} \bre
&+ (\R)_{pi} z_{ia} (\Lg)_{aq} +(\R)_{pa} z_{ai} (\Lg)_{iq}.
\end{align}
One can assure that Eq.(\ref{eq:Lx}) is consistent with SUHF for $|\Psi\ket = |\Phi\ket$ and ${\bf z} = {\bf 0}$.

Comparing the terms in Eq.(\ref{eq:Lx}), one can easily identify the spin-incomplete relaxed 1PDM as the sum of the unrelaxed density matrix and relaxation correction (integrated),
%\begin{align}
%P^{\rm rel}_{\nu\mu} = \bra \Psi|a_\mu^\dag a_\nu \hat P |\Psi\ket + \sum_g w_g n_g \Bigl[\tcalR_{\nu a} z_{ai} \tcalL_{i \mu} +  \tcalR_{\nu i} z_{ai} \tcalL_{a \mu} +\rho_{\nu\mu} (\calW_{ia} +\calW_{ai} -N_{ia}-N_{ai} ) z_{ai}\Bigr] \label{eq:Prel}
%\end{align}
%in the AO basis, or it can be shown (see Appendix) that it is elegantly given in the MO basis
\begin{align}
P^{\rm rel}_{pq} = \sum_g w_g n_g ({\bf P}_g + {\bf P}_g^{\rm corr})_{pq}.\label{eq:Prel}
\end{align} 
To gain some    physical insights, one can elegantly rewrite the integrated correction term as
\begin{align}
 \sum_g w_g n_g ({\bf P}_g^{\rm corr})_{pq} =   z_{ai}  \frac{\dd \bra \Phi| a_q^\dag a_p \hat P |\Phi\ket}{\dd \kappa_{ai}^*} + z_{ia} \frac{\dd \bra \Phi | a_q^\dag a_p \hat P |\Phi\ket }{\dd \kappa_{ai}}.\label{eq:Prel2}
\end{align} 
The meaning of Eq.(\ref{eq:Prel2}) is striking; it accounts for the first-order orbital relaxation effect on the ECISD density matrix via the reference (SUHF) density matrix.
After some simple algebra, one can verify that the Wigner-Eckart theorem can be directly applied so as to obtain the spin-adapted relaxed density; in other words, ${\bf D}^{\rm rel}_{\alpha}$ and ${\bf D}^{\rm rel}_\beta$ can be derived from a linear combination of ${\bf P}^{\rm rel}_{\alpha\alpha}, {\bf P}^{\rm rel}_{\beta\alpha}, {\bf P}^{\rm rel}_{\alpha\beta},$ and ${\bf P}^{\rm rel}_{\beta\beta}$ (see Appendix). 

Similarly, the relaxed 2PDM can be easily derived, and may be explicitly symmetrized in the AO basis without loss of generality, i.e., $\tilde P_{\mu\nu,\lambda\sigma} = \frac{1}{4} {\mathscr P}_{\mu\nu} {\mathscr P}_{\lambda\sigma}P_{\mu\nu,\lambda\sigma}$. However, for ease of computations, the non-separable (unrelaxed) term should be directly contracted with two-electron integrals in the AO basis to avoid prohibitively large memory requirement and disk storages.

%Finally, note that the  $\hat P \rightarrow 1$ limit correctly reproduces standard relaxed density because $\tcalL_{p \mu} \rightarrow C^*_{\mu p}$, $\tcalR_{\mu p} \rightarrow C_{\mu p}$, and $\calW_{pq}, N_{pq} \rightarrow \delta_{pq}$. 

\subsection{ECISD+Q gradient}\label{sec:ECISD+Q}
We should mention the gradient of the Davidson correction,\cite{Langhoff74,Davidson74,Tsuchimochi16A} $\Delta E_{\rm Q}$, can be formulated. In this case, the total energy $E+\Delta E_{\rm Q}$ is stationary with respect neither to {\bf C} nor to ${\bf c}$. Hence, our Lagrangian takes a more complicated form
\begin{align}
{\mathscr L}_{\rm Q} &= E+ \Delta E_{\rm Q} + \sum_{ai} \left(\calF_{ai} +\calF_{ia}\right) z_{ai} -\Tr \left[{\bm\epsilon}\left({\bf C}{\bf S}^\dag{\bf C} - {\bf 1}\right)\right]  \bre
&+ \sum_I \tilde z_I \left(\bra \Phi_I | (\hat H - E) \hat P |\Psi\ket + \bra \Psi |(\hat H - E) \hat P| \Phi_I\ket \right) 
\end{align}
where $\Phi_I \in \Phi, \Phi_i^a, \Phi_{ij}^{ab}$, requiring to solve the resulting CPECISD equation, $\dd {\mathscr L}_{\rm Q}/\dd {\bf c} = {\bf 0}$, for $\tilde {\bf z}$. While it is straightforward to derive the equations, this will significantly complicate the algorithm and thus is beyond the scope of the present work. 

\section{Numerical examples}\label{sec:Results}
\subsection{Computational details}
The above scheme has been implemented in the GELLAN suite of programs, which can also handle the analytical gradient of SUHF. All of the calculations were performed with all-electrons correlated unless otherwise specified. While incorporating the frozen-core approximation is possible, the implementation is not as straightforward as that for the conventional post-HF methods, because ECISD requires a constrained SCF optimization to define frozen-core orbitals.\cite{Tsuchimochi16B,Tsuchimochi10,Tsuchimochi11} We will address this issue elsewhere.

For the number of grid points for the spin-projection, $N_{\rm grid}= 4$ was found to be enough to make $\bra \hat S^2 \ket$ of a SUHF wave function precise to at least $10^{-9}$  for all the calculations presented in this work.
If a precision of $10^{-7}$ is requested for $\bra \hat S^2\ket$ as in the previous studies,\cite{Schutski14} only three grid points is sufficient, while the resulting energy error is less than 1 $\mu E_{\rm h}$. Also, increasing $N_{\rm grid}$ did not change both the SUHF and ECISD wave functions. All the spin-projected calculations employed GELLAN, and single reference methods including  CC singles and doubles (CCSD) and CCSD with perturbative triples, CCSD(T), were performed using {\sc Gaussian}.\cite{G09}
We also used {\sc Molpro}\cite{Molpro} for the MR calculations.

\subsection{Ozone}\label{sec:ozone}
We take the ozone molecule as our first example, as it has been extensively studied by other authors due to its degenerate electronic structure.\cite{Lee87,Borowski92,Li99, Kowalski05,Hino06,Kalemos08} We use Dunning's DZP\cite{Hay75} basis set following the earlier work of reduced MR-CCSD (RMR-CCSD) calculations,\cite{Li99} for a direct comparison with our results.

\begin{table*}
\caption{Total energy and geometry of ozone computed with the DZP basis.}\label{tb:O3}
\tabcolsep = 2mm
\begin{threeparttable}[t]
\begin{tabular}{lcccccccc}
\hline\hline
Method &Energy (a.u.) & $R_{\rm OO} ({\rm \AA})$&$\angle$OOO ($^\circ$)  & \multicolumn{3}{ c}{  $\omega$ (cm$^{-1}$) } & $\mu$ ($D$) \\
\cline{5-7}
&&&&\;\;\; $1a_1$  \;\;\;&  \;\;\;$2a_1$ \;\;\;&  \;\;\;$1b_2$ \;\;\;  \\
\hline 
RHF & --224.320 897 & 1.207 & 118.9  &1541 &  842 & 1432& 0.874 \\
SUHF &  --224.438 884 & 1.284  & 114.4  & 722 & 936 & 226 & 0.191  \\
CASSCF(2,2) &   --224.403 040 &  1.258 & 115.1  &  1181 & 776 & 1492& 0.220 \\ 
RCISD &--224.858 121 & 1.247 & 117.7 & 1388 & 783 & 1562 & 0.702 \\
ECISD & --224.898 361 & 1.274 & 116.2  &1209 & 735&1349 & 0.439\tnote{a}\\

MRCISD(2,2) &--224.890 282 & 1.271 & 116.2 & 1226 & 746 & 1358  &  0.347\tnote{b} \\
RCCSD & --224.943 339 & 1.275& 117.1 & 1249 & 729 & 1244& 0.595  \\
RMR-CCSD\tnote{c} &--- & 1.277 & 116.7 & 1187 & 727 & 1156 & --- \\
Exp. & --- &1.272 & 116.8 & 1135 & 716 & 1089 & 0.532 \\
\hline\hline
\end{tabular}  
{\footnotesize
\begin{tablenotes}
\item[a] 0.355 for unrelaxed density.
\item[b] Unrelaxed.
\item[c] $1s$ orbitals are frozen and basis exponent $\alpha_D$ is modified to 1.211.
\end{tablenotes}
}
\end{threeparttable}
\end{table*}

We have performed geometry optimizations on this system with various methods including RHF, SUHF, ECISD, CISD, CCSD, and MRCISD, and the results are listed in Table \ref{tb:O3}. The results of reduced MR-CCSD  are taken from Ref.[\onlinecite{Li99}] for comparison. All the MR methods employ the minimum (2$e$,2$o$) active space. Using the finite difference of analytical gradients, we also evaluated the vibrational frequencies. RHF significantly underestimates the bond length due to the lack of static correlation, while SUHF, in spite of its mean-field  nature, gains a large amount of correlation energy, predicting a much better geometry. It is noteworthy that SUHF outperforms CASSCF for the energy and geometry. However, it turned out that SUHF fails to predict correct frequencies even qualitatively for this simple molecule.
Especially, the $1b_2$ mode becomes unreasonably small compared to the other methods. Although the frequency for the $2a_1$ mode is overestimated, that for $1a_1$ results in a too low value. This indicates that SUHF's potential energy surface is unphysically shallow for this particular system, which is somewhat astonishing, given  its good performance on the energy and geometry. We note that UHF, without spin-projection, provides more reasonable frequencies, and so does the non-collinear spin-projection on generalized HF (although not listed in the table).\cite{Roman} Hence, this is an SUHF-specific failure and a precaution is required when SUHF is used for frequency calculations.

\begin{figure}
\includegraphics[width=75mm, bb = 0 0 2284 2855]{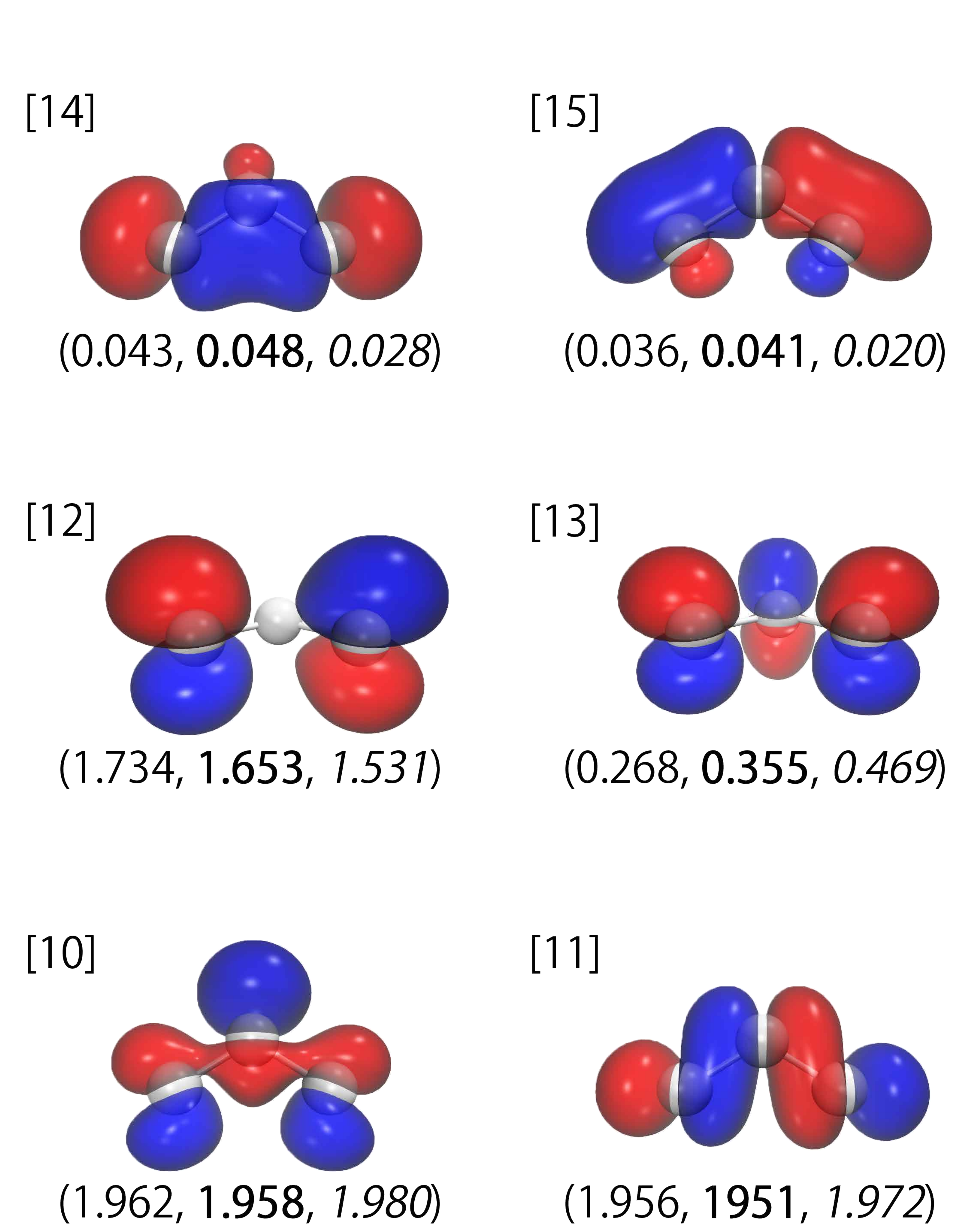}
\caption{Selected natural orbitals of O$_3$. All the orbitals look qualitatively similar for ECISD and SUHF. The corresponding natural occupations are listed in the following order; ECISD (relaxed), ECISD (unrelaxed) in bold, and SUHF in italic.} \label{fig:NO}
\end{figure}

On the other hand, ECISD certainly improves the SUHF results in all of the aspects.
Not only does it correct the ill-behaved frequencies of SUHF, but also its energy and geometry are comparable to those of MRCISD.
Employing the same active space as MRCISD, RMR-CCSD delivers as accurate results especially for vibrational frequencies.
Despite its single-reference nature, CCSD also  shows an excellent performance,  yet ameliorating RHF only partially. Hence, we conjecture that the discrepancies between the ECISD and MRCISD frequencies and those of RMR-CCSD as well as experimental values are mostly attributed to the lack of linked disconnected terms in the former, although  the use of larger basis sets could change the results greatly. 

We also computed the dipole moment $\mu$ with the unrelaxed and relaxed densities of ECISD. Without the orbital relaxation effect, ECISD gives 0.355 Debye ($D$); a solid improvement is obtained over SUHF (0.191 $D$) and CASSCF (0.220 $D$). We should stress that this value is also comparable to  the MRCISD result (0.347 $D$), computed with its unrelaxed density. It is noteworthy that the underestimation of the unrelaxed density on $\mu$ is further improved by the response correction, whose contribution is substantial, giving a total $\mu$ of 0.439 $D$. 

To see the main difference between SUHF and ECISD relaxed and unrelaxed densities, in Figure \ref{fig:NO}, we have visualized the NOs with their occupation numbers. For comparison, here we use the ECISD geometry also for the SUHF results, which yield $\mu = 0.195 D$. 
The appearances of NOs computed with the relaxed and unrelaxed ECISD densities and SUHF density are indistinguishable with one another, so we only depict the relaxed ECISD orbitals.
It is worth mentioning that, while the most orbitals have almost the same occupancies for the relaxed and unrelaxed densities, the degenerate orbitals are more sensitive to the orbital relaxation effect, i.e., the occupation numbers of 12th (non-bonding) and 13th (anti-bonding) NOs noticeably change. The relaxed occupation of the latter is smaller than that of the unrelaxed calculation, reflecting the more ``dynamical'' character of the wave function. This is responsible for the description of $\mu$, because the density is more polarized without occupying electrons in the anti-bonding NO as clearly seen in Figure \ref{fig:NO}. 
The non-bonding and anti-bonding orbital occupations of SUHF are even more fractional than the unrelaxed ECISD density due to the neglect of the vast majority of dynamical correlation effects. The SUHF density becomes even less polarized by promoting more electrons than necessary, from the non-bonding NO to the anti-bonding one.

\begin{figure}[t!]
\includegraphics[width=80mm, bb = 0 0 2000 1450]{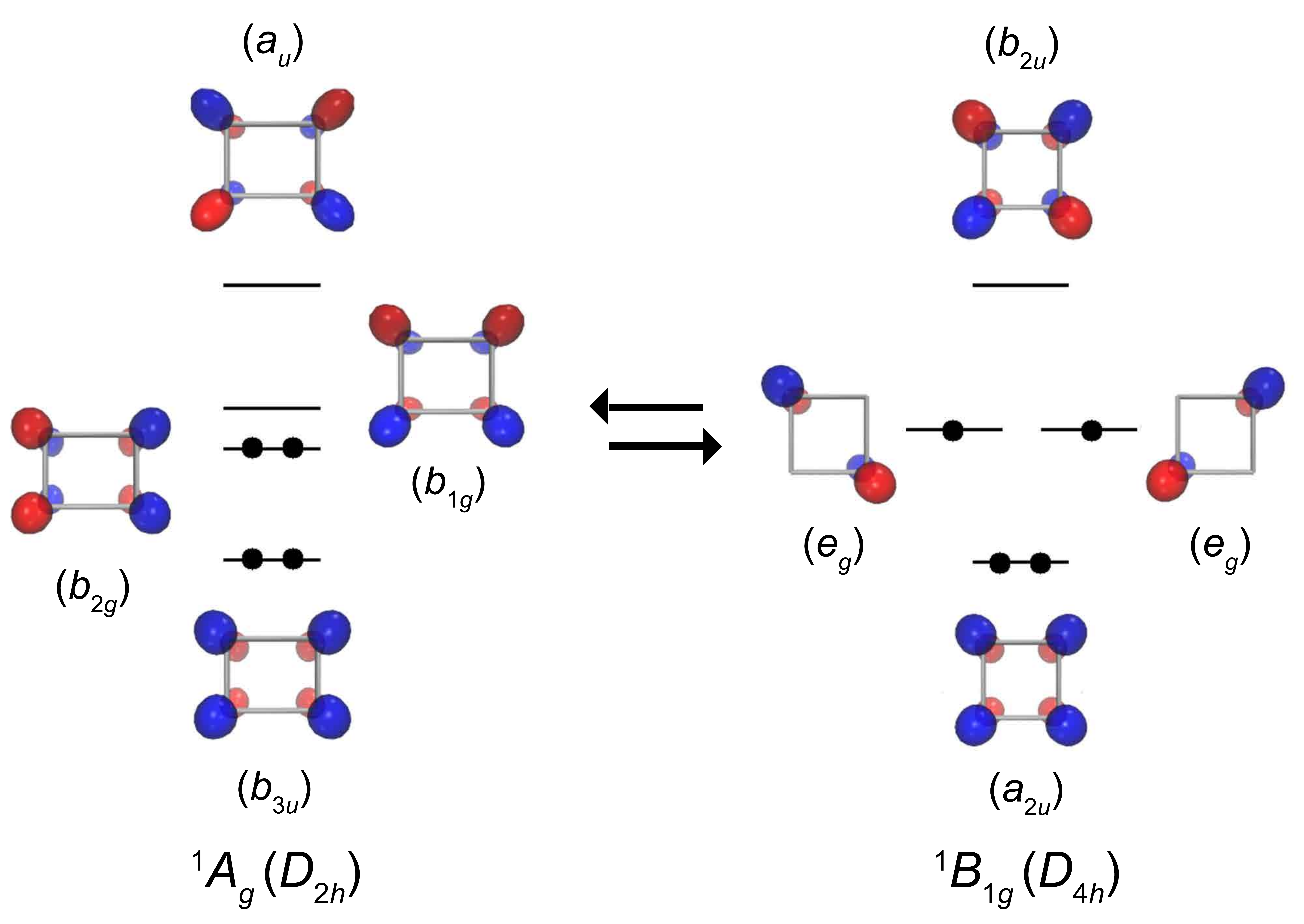}
\caption{Automerization of singlet cyclobutadiene.} \label{fig:D24h}
\end{figure} 

\subsection{Cyclobutadiene} 
Cyclobutadiene is another suitable test case that exhibits a degenerate electronic structure. 
Although this molecule is highly distabilized due to the anti-aromaticity of $4n\pi$ electrons, the characterization of its automerization has posed a significant challenge and thus gained considerable attention both experimentally\cite{Chapman73A,Chapman73B,Whitman82,Carpenter83} and theoretically.\cite{Balkova94,Levchenko04,Eckert06,Bhaskaran08,Li09,Shen12} 
As depicted in Figure \ref{fig:D24h}, the $^1A_g$ ground state has the distorted rectangular equilibrium geometry because of the Jahn-Teller effect, and most single-reference methods should offer a reasonable vertical singlet-triplet gap $\Delta E_{\rm ST}$ as the static correlation is not strong enough. However, at the transition structure (square $D_{4h}$) of the automerization, its electronic structure is entangled, requiring MR treatments or explicit inclusion of triple excitations such as CCSDT. Therefore, single-reference methods limited to doubles typically overestimate the the activation barrier $\Delta E^\ddag$ and significantly underestimate the singlet-triplet gap at the square structure $\Delta E_{\rm ST}^*$. Balkov\'a and Bartlett studied this system using MR-CCSD(T) with a small DZP basis and frozen orbitals, and reported $\Delta E^\ddag = 6.6$ kcal/mol and $\Delta E_{\rm ST}^* = 6.9$ kcal/mol.\cite{Balkova94} On the other hand, Levchenko and Krylov used the larger cc-pVTZ basis functions to perform all-electron equation-of-motion SF-CCSD (EOM-SF-CCSD), which yielded 7.5 and 8.5 kcal/mol for $\Delta E^\ddag$ and $\Delta E^*_{\rm ST}$.\cite{Levchenko04} Many other computational studies focused on $\Delta E^\ddag$ but not on $\Delta E_{\rm ST}^*$. Here we will perform geometry optimizations to calculate all these quantities using ECISD and MRCI.

\begin{figure}[t!]
\includegraphics[width=80mm, bb = 0 0 195 127]{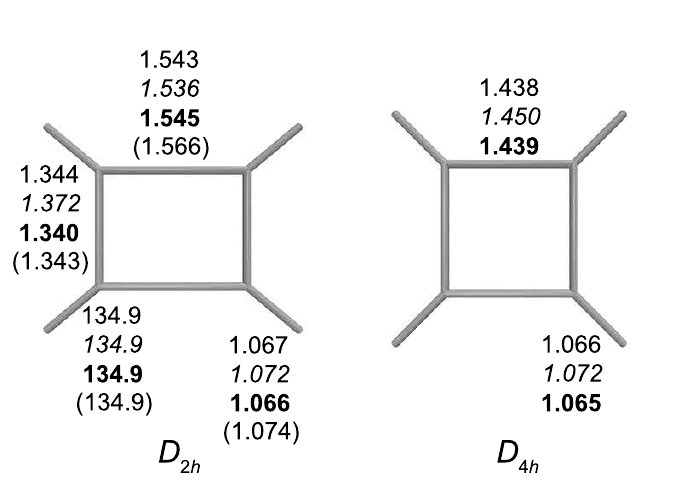}
\caption{Optimized geometries of the ground state singlet states computed with ECISD, SUHF (italic), and MRCI (bold) in $\rm\AA$ and degree. For $D_{2h}$, CCSD(T) results are also shown in parentheses.} \label{fig:Geo}
\end{figure}

Figure \ref{fig:Geo} shows the optimized geometries of the rectangular and square structures computed with ECISD and SUHF (in italic). For comparison, we list the result of MRCI in bold, where we used an active space of $(4e,4o)$ that consists of the orbitals shown in Figure \ref{fig:D24h}. For $D_{2h}$, CCSD(T) results are also provided in parentheses. While SUHF geometries are reasonable, the agreement between ECISD and MRCI geometries is excellent. Their ground state energies are also close to each other, as tabulated in Table \ref{tb:C4H4}, indicating their wave functions are similar both qualitatively and quantitatively. However, it turns out that both ECISD and MRCI predict slightly shorter bond lengths for $D_{2h}$ compared to CCSD(T). This is partly attributed to the fact that both ECISD and MRCI are size-inextensive. In addition, CCSD(T) also has its own deficiency; the method is not suitable for describing static correlation, which is found even in $D_{2h}$. Although CCSD(T) performs very well compared to MR-CCSD(T) at this structure as shown by Balkov\'a and Bartlett (the energy difference is less than 2 mH at a DZP basis), the optimized geometry can be largely affected by the presence of static correlation; for example, MR-CCSD predicted 1.570 and 1.367 ${\rm \AA}$ for $R_{\rm CC}$, while CCSD gave 1.582 and 1.359 ${\rm \AA}$.\cite{Balkova94}

\begin{figure}[t!]
\includegraphics[width=80mm]{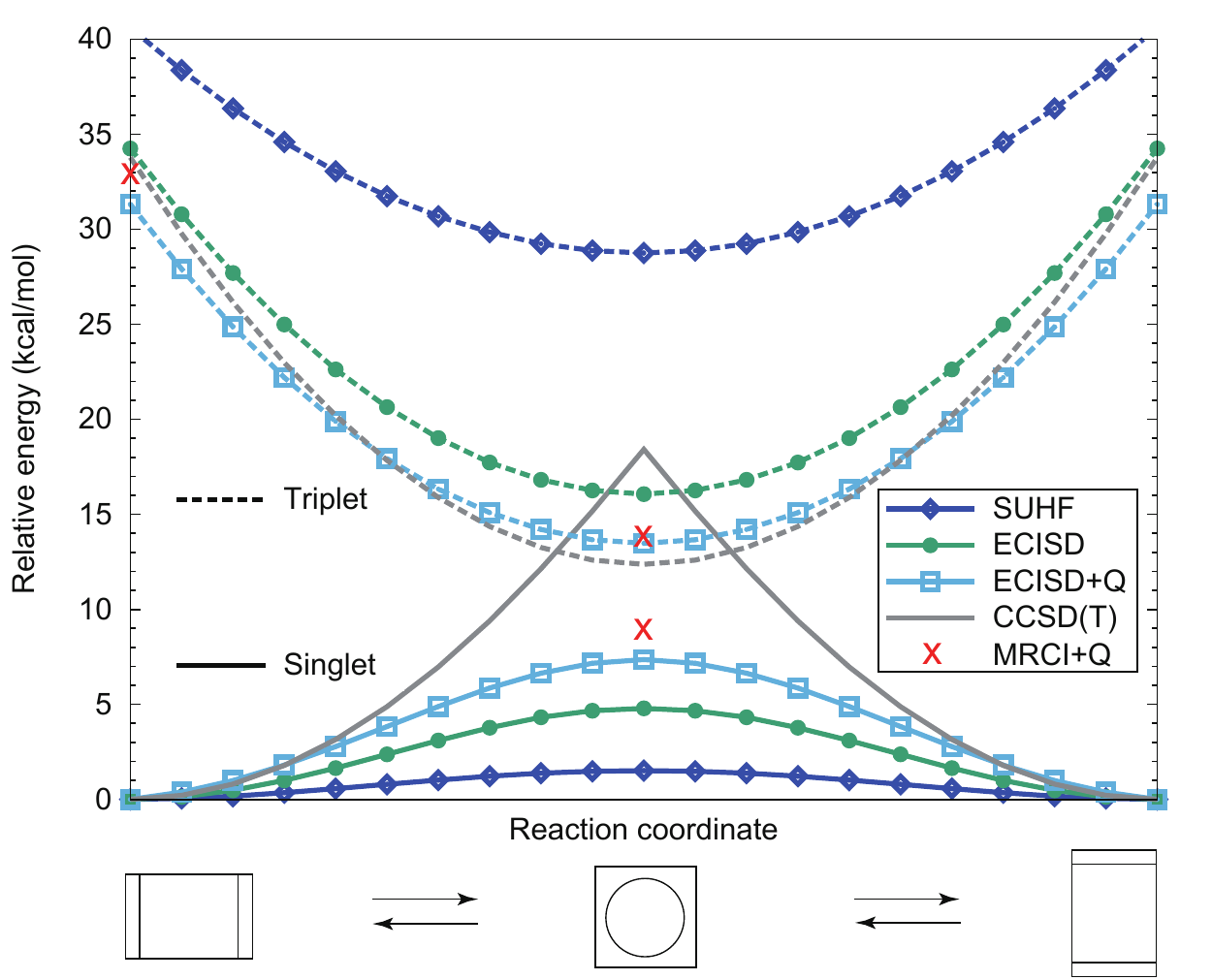}
\caption{Energy profiles along with the automerization pathway. The solid and broken curves represent singlet and triplet states, respectively.} \label{fig:C4H4}
\end{figure}

Nonetheless, these optimized geometries allow us to draw energy diagrams for the automerization at each level of theory. Figure \ref{fig:C4H4} presents the lowest singlet and triplet potential curves (solid and broken curves, respectively) along with the reaction coordinate that is defined as a linear interpolation between the two structures. We choose the low-spin triplet state for spin-projection in this study, because the low-spin SUHF energy was found to be much lower than the high-spin one (the stability gained in the former is 35 mH compared to the latter at $D_{2h}$), suggesting the former is a qualitatively better starting point for subsequent ECISD calculations. Also, the use of low-spin states was recommended for the calculations of singlet-triplet gaps in our previous work.\cite{Tsuchimochi16B, Tsuchimochi17B} Note that, nevertheless, the ECISD+Q results for $\Delta E_{\rm ST}$ and  $\Delta E_{\rm ST}^*$ in this case did not significantly change when a high-spin state was used; the largest difference between high-spin and low-spin triplet energies is less than 3 mH with a non-parallelity error of about 0.1 mH.

The activation barrier of SUHF is found to be too low with only 1.5 kcal/mol, which is consistent with the finding in Section \ref{sec:ozone} that the SUHF energy potential sufraces appear to be unphysically shallow. Inclusion of singles and doubles as well as the Davidson correction due to Pople {\it et al.}\cite{Pople77,Duch94} increases $\Delta E^\ddag$ to 4.8 and 7.4 kcal/mol, the latter almost coinsiding the result of EOM-SF-CCSD (7.5 kcal/mol). However, as tabulated in Table \ref{tb:C4H4}, the ECISD+Q value is lower than that of MRCI+Q by 1.7 kcal/mol. Nevertheless, we note that the correlation energy gained by ECISD is comparable to that of MRCI at $D_{2h}$, while the former is much larger than the latter at $D_{4h}$, as can be seen from the large difference in the computed barrier heights. This suggests that ECISD spans a better reference space than does MRCI (for singlet) for the subsequent Davidson correction, and therefore, we conclude the singlet ECISD+Q results should be more reliable than those of MRCI+Q with this active space.

\begin{table}
\tabcolsep = 2mm
\begin{threeparttable}[t]
\caption{Total energy of ground state at the $D_{2h}$ geometry ($a.u.$), activation barrier, and vertical singlet-triplet gaps at both $D_{2h}$ and $D_{4h}$ geometries (kcal/mol). }\label{tb:C4H4}
\begin{tabular}{lcrcc}
\hline\hline
%Method & Energy ($D_{2h}$) & $\Delta E^\ddag$  &   \multicolumn{2}{c}{$\Delta E_{\rm ST}$ ($D_{2h}$)\tnote{a}}  & \multicolumn{2}{c}{ $\Delta E_{\rm ST}$ ($D_{4h}$)\tnote{a}} \\
Method & Energy ($D_{2h}$) & $\Delta E^\ddag$  &   $\Delta E_{\rm ST}$ & $\Delta E_{\rm ST}^*$ \\\hline
SUHF & --153.790 05 & 1.5 & 40.6 &27.2       \\ % (T0)
ECISD & --154.344 81 & 4.8 & 34.3 & 11.3       \\ % (T0)
ECISD+Q\tnote{a} & --154.431 67 & 7.4 & 31.3 & \:\:6.1\\ % (PC T0)
%ECISD+Q\tnote{a} & --154.431 67 & 7.4 & 33.1& \:\:7.8        \\ % (PC T1)
MRCI\tnote{b} & --154.341 53 & 8.4 & 34.5&     \:\:6.3       \\ 
MRCI+Q\tnote{b,c} & --154.433 56&  9.1 & 33.0 & \:\:4.7       \\ 
CCSD(T)\tnote{d,e}& --154.453 26  &18.4 & 33.8 & --6.1       \\
EOM-SF-CCSD\tnote{e,f} & --154.424 95 & 7.5 & 38.3 & \:\:8.5       \\
%EOM-SF-CCSD\tnote{g} & --154.442 54 & 18.3 & 38.3 & 8.5  &     \\
 \hline\hline
\end{tabular} 
{\footnotesize
\begin{tablenotes}
%\item[a] Values in parentheses are computed with $m_s = 0$ low-spin triplet states.
%\item[a] The SUHF geometries.
%\item[b] The ECISD geometries.
\item[a] ECISD geometries are used.
\item[b] The active space used is $(4e,4o)$.
\item[c] MRCI geometries are used.
\item[d] RHF/ROHF reference. 
\item [e] The singlet ($D_{2h}$) and triplet ($D_{4h}$) CCSD(T) geometries are used.
\item[f] UHF reference. Taken from Ref.[\onlinecite{Levchenko04}].
%\item[g]  ROHF reference with the triplet CCSD(T) geometry. Taken from Ref.[XXX].
\end{tablenotes}
}
\end{threeparttable}
\end{table}

On the other hand, CCSD(T) produces a kink at the $D_{4h}$ point as expected, because of the exact degeneracy in the $e_g$ orbitals.
As a result, its activation barrier is predicted to be 18.4 kcal/mol, which is significantly larger than the experimental estimate of 1.6--10 kcal/mol,\cite{Whitman82}
resulting in the incorrect ordering of the singlet and triplet energies.
%{\red and the energetic resulting in the incorrect ordering of the singlet and triplet states is flipped energies}.
However, again, the small energy difference between CCSD(T) and MR-CCSD(T) in Ref.[\onlinecite{Balkova94}] validates the accuracy of CCSD(T) at the $D_{2h}$ geometry. Hence its $\Delta E_{\rm ST}$, 33.8 kcal/mol, is expected to be reliable, which is indeed close to the MRCI+Q result (33.0 kcal/mol). While EOM-SF-CCSD  gives a somewhat larger value of 38.3 kcal/mol, the ECISD+Q prediction agrees well with CCSD(T) and  MRCI+Q. We observe a similar behavior for $\Delta E_{\rm ST}^*$, where the EOM-SF-CCSD value is again slightly larger than those of ECISD+Q and MRCI+Q. It is likely that this discrepancy is attributed to the spin-contamination inevitable in SF calculations.\cite{Sears03,Tsuchimochi15C} At the $D_{4h}$ structure, ECISD+Q and MRCI+Q are energetically very close to each other; the singlet energies of them are -154.41992 and -154.41910, whereas the triplet energies are -154.41018 and -154.41166, in $E_{\rm h}$. Therefore, ECISD+Q slightly outperforms MRCI+Q for singlet, but gets worse for triplet, resulting in the increase of $\Delta E_{\rm ST}^*$ compared the latter, although small (1.4 kcal/mol). 

Overall, ECISD produces similar results to those obtained with MRCI, especially when a size-consistency correction is introduced. It appears ECISD is satisfactory for this system, considering the simplicity of its wave function {\it ansatz}. 

%%%%%%
% TABLE %
%%%%%%

%\subsection{Benzyne}
%Next, we investigate benzyne structures, namely $o$-, $m$-, and $p$-benzynes. In these systems, the ground state is singlet and exhibits an open-shell electronic structure. Especially, $m$-benzyne is an enlightening example because CCSD predicts that the molecule has two rings instead of one. This problem was studied by Paldus, and only the inclusion of perturbative triples fixes it. Also, in the previous work, Schutski {et al.} reported SUHF successfully captures the diradical characters of these molecules, but it breaks the $C_{2v}$ spatial symmetry of $m$-benzyne and predicts $C_s$ to be more stable. We also performed the SUHF calculations with spherical harmonics and reproduced the same problem for $m$-benzyne. Therefore, it is interesting if ECISD can fix these issues.

%We have used a 6-311G(d,p) basis to perform the geometry optimization on these systems.  

\section{Conclusions}\label{sec:Conclusions}
The availability of analytical derivative is critical  in electronic structure calculations, as it broadens the application range of a method by providing a means to compute molecular properties and nuclear gradients. In this manuscript, we have demonstrated that, using the Z-vector technique, the first derivative of post-SUHF methods can be derived in a manner analogous to  single-reference post-HF. The chief difference is that, in the MO basis, the density matrices of spin-projected approaches explicitly depend on the underlying molecular orbitals and AO overlap matrix, while only particle-hole amplitudes such as CI coefficients are relevant in single-reference cases. We showed the calculation of density matrix derivatives with respect to nuclear coordinates can be avoided by expressing the total energy as a sole function of contractions in terms of the nonorthogonal Wick theorem, which allows for complete cancellation in such terms. The resulting relaxed density matrices are not spin-adapted, if a projection operator is not explicitly present at both bra and ket states. However, since the relaxation correction is written as the gradient of SUHF 1PDM with respect to orbital change, one can retrieve the spin-adapted form using the Wigner-Eckart theorem.

Our results revealed the SUHF frequencies can be sometimes even much worse than UHF as seen in ozone, but this deficiency can be largely ameliorated by the introduction of dynamic correlation in ECISD. The dipole moment of ozone was also greatly improved when computed with the relaxed density. For the automerization of cyclobutadiene, the ECISD+Q results agreed quite well with MRCI, as well as with EOM-SF-CCSD and CCSD(T) reference values. However, our results also strongly indicate the inclusion of size-extensive (consistent) correction is indispensable for accurate descriptions of these systems. While the Davidson correction greatly improves the energy with a negligible computational cost, its derivative is evidently cumbersome. Hence it is more advantageous to renormalize the effect of $\Delta E_{\rm Q}$ into the energy functional, so that the total energy derivative with respect to ${\bf c}_I$ remains to be zero. Research for such extension is currently underway; some preliminary results are reported elsewhere.\cite{Tsuchimochi17B}

\section*{Acknowledgements}
We would like to thank Motoyuki Uejima for helpful discussions, and Roman Schutski for performing non-collinear projected HF calculations. This work was supported by MEXT's FLAGSHIP2020 as priority issue 5 (development of new fundamental technologies for high-efficiency energy creation, conversion/storage and use). We are also indebted to the HPCI System Research project for the use of computer resources (Project ID: hp150228, hp150278).

\section*{Appendix}
\appendix
\section{Wigner-Eckart theorem for relaxed density matrix}
The relaxed density matrix in the form of Eq.(\ref{eq:Prel}), ${\bf P}^{\rm rel}$, is not spin-adapted because the perturbation has been applied to the half-projected Hamiltonian $\hat H(\lambda)\hat P$ but not to the explicitly projected one $\hat P \hat H(\lambda)\hat P$. 
On the contrary, it can be even asymmetric. This section shows the Wigner-Eckart theorem can be equally applicable to derive ${\bf D}^{\rm rel}$ from  ${\bf P}^{\rm rel}$. 

For our purpose, let us split the spin-adapted relaxed density matrix as
\begin{align}
{\bf D}^{\rm rel} = {\bf D} + {\bf D}^{(vo)} + {\bf D}^{(ov)},
\end{align}
where the first term is the unrelaxed spin-adapted ECISD density matrix ($\bra \Psi |\hat P a_p^\dag a_q \hat P |\Psi\ket$), and superscripts $(vo)$ and $(ov)$ stand for the contributions from the corresponding space $z_{ai}$ and $z_{ia}$, respectively. 

For the explicitly spin-projected Hamiltonian, $\hat P \hat H \hat P = \sum_{g,g'} w_g w_{g'} \hat R_{g'} \hat H \hat R_g$, it can be shown that PHF Fock for the $ai$ component ($\kappa=\alpha,\beta$ spin) is
\begin{widetext}
\begin{align}
\bar {\bar \calF}_{a\kappa,i\kappa}& =\frac{1}{\bra \Phi|\hat P |\Phi\ket} \sum_{g,g'} w_g w_{g'} n_{g'g} \Biggl((\bar {\bar E}_{g'g}-\bar {\bar E}_{\rm PHF})  \frac{\bra \Phi | a_{i\kappa}^\dag a_{a\kappa} \hat R_{g'} \hat R_g |\Phi\ket }{n_{g'g}} + ({\bar{\bar{\bf F}}}_{g' g})_{p\sigma,q\sigma} \frac{\bra \Phi | a_{a\kappa} \hat R_{g'} a_{p\sigma}^\dag \hat R_g |\Phi\ket }{n_{g'g}}   \frac{\bra \Phi | a^\dag_{i\kappa} \hat R_{g'} a_{q\sigma} \hat R_g |\Phi\ket }{n_{g'g}} 
\Biggr),
\end{align} 
where
\begin{align}
n_{g'g} &= \bra \Phi| \hat R_{g'} \hat R_g |\Phi\ket,\\
\bar {\bar E}_{g'g}& = \sum_{pq} \sum_{\sigma} h_{pq}  \frac{\bra \Phi |  \hat R_{g'} a_{p\sigma}^\dag  a_{q\sigma} \hat R_g |\Phi\ket }{n_{g'g}}  +\frac{1}{2} \sum_{pqrs}\sum_{\sigma\kappa} \bra pq||rs \ket  \frac{\bra \Phi |  \hat R_{g'} a_{p\sigma}^\dag  a_{r\sigma} \hat R_g |\Phi\ket }{n_{g'g}}  \frac{\bra \Phi |  \hat R_{g'} a_{q\kappa}^\dag  a_{s\kappa} \hat R_g |\Phi\ket }{n_{g'g}}\\
\bar {\bar E}_{\rm PHF} &= \frac{1}{\bra \Phi| \hat P |\Phi\ket } \sum_{g,g'} w_g w_{g'} n_{g'g} \bar {\bar E}_{g'g},\\
({\bar{\bar{\bf F}}}_{g' g})_{p\sigma,q\sigma} &=h_{p\sigma,q\sigma} + \sum_{rs,\kappa}\bra pr||qs\ket  \frac{\bra \Phi |  \hat R_{g'} a_{r\kappa}^\dag  a_{s\kappa} \hat R_g |\Phi\ket }{n_{g'g}}.
\end{align} 

Note that the spin-adapted relaxed density contribution in the MO basis $D^{(vo)}_{q\sigma,p\sigma}$ then comes from $\bar {\bar E}_{g'g}^{(\bar x)}$, $\bar {\bar E}_{\rm PHF}^{(\bar x)}$, and $({\bar{\bar{\bf F}}}_{g' g}^{(\bar x)})_{p\sigma_p,q\sigma_q}$. Therefore, we have
\begin{align}
D^{(vo)}_{q\sigma,p\sigma} &= \sum_\kappa z_{a\kappa,i\kappa}\sum_{gg'} w_g w_g' n_{g'g} \Biggl( \frac{\bra \Phi | a_{a\kappa} \hat R_{g'} a_{p\sigma}^\dag \hat R_g |\Phi\ket }{n_{g'g}} \frac{\bra \Phi | a^\dag_{i\kappa} \hat R_{g'} a_{q\sigma} \hat R_g |\Phi\ket }{n_{g'g}} + \frac{\bra \Phi |\hat R_{g'} a_{p\sigma}^\dag a_{q\sigma} \hat R_g |\Phi\ket }{n_{g'g}} \frac{\bra \Phi |  a^\dag_{i\kappa} a_{a\kappa} \hat R_{g'} \hat R_g |\Phi\ket }{n_{g'g}} \bre
&- \frac{\bra \Phi |\hat R_{g'} a_{p\sigma}^\dag a_{q\sigma} \hat R_g |\Phi\ket }{n_{g'g}} \sum_{G'G} w_G w_{G'} n_{G'G} \frac{\bra \Phi |  a^\dag_{i\kappa} a_{a\kappa} \hat R_{G'} \hat R_G |\Phi\ket }{n_{G'G}}  \black \Biggr).
\end{align}
\end{widetext}
The last term is simply a product of two projected elements; both $n_{g'g}$ and $n_{G'G}$ cancel out between the numerator and denominator, and one can use the definition $\hat P \equiv \sum_g w_g \hat R_g$. On the other hand, $n_{g'g}$ in the the first two terms also cancel out by using the nonorthogonal Wick theorem in a backward manner, and these terms can be cast as a single term. Finally, we find
\begin{widetext}
\begin{align}
D^{(vo)}_{q\sigma,p\sigma} &= \sum_\kappa z_{a\kappa,i\kappa}\Biggl(\sum_{gg'} w_g w_g' n_{g'g}  \frac{\bra \Phi | a_{i\kappa}^\dag a_{a\kappa} \hat R_{g'} a_{p\sigma}^\dag a_{q\sigma} \hat R_g |\Phi\ket }{n_{g'g}}\black
-  \bra \Phi| a_{i\kappa}^\dag a_{a\kappa} \hat P |\Phi\ket  \bra \Phi |\hat P a_{p\sigma}^\dag a_{q\sigma} \hat P |\Phi\ket   \black \Biggr)\bre
 &= \sum_\kappa z_{a\kappa,i\kappa}  \Biggl(\bra \Phi | a_{i\kappa}^\dag a_{a\kappa} \hat P a_{p\sigma}^\dag a_{q\sigma} \hat P |\Phi\ket 
 -   \bra \Phi| a_{i\kappa}^\dag a_{a\kappa} \hat P |\Phi\ket \bra \Phi |\hat P a_{p\sigma}^\dag a_{q\sigma} \hat P |\Phi\ket  \Biggr),
\end{align}
\end{widetext}
for which we can resort to the Wigner-Eckart theorem for the expansion of $\hat P a_{p\sigma}^\dag a_{q\sigma} \hat P$ to a linear combination of half-projected, mixed-spin operators $a^\dag_{p\lambda} a_{q\tau} \hat P$.

\ifx11
\section{Alternative derivation of energy-weighted density matrix in PHF}
Here we check if the energy-weighted density matrix {\bf W} in PHF can also be obtained by only considering the difference between $\tilde {\bm \calL}_g^{(x)}$ and $\dd\tilde {\bm \calL}_g/\dd \kappa^*_{pq}$, that is, Eq.(\ref{eq:x_and_kappa}). Again, for the $\kappa^*_{pq}$ differentiation, we have to write the PHF energy as a function of only ${\bm\calW}_g$, $\tilde{\bm\calL}_g$, and $\tilde{\bm\calR}_g$ to assure no explicit ${\bC}^\dag$ dependence, so that one can use the chain-rule as in Eq.(\ref{eq:Fock1}). For PHF, it is evident that ${\bf z} \rightarrow {\bf 0}$ and $\calE_g \rightarrow E_g$, so the energy-weighted density matrix {\bf W} is defined as
\begin{align}
W_{pq} =  -  C_{\mu p}^* \sum_g w_g n_g \frac{\dd E_g}{\dd  (\tilde{\bm\calL}_g)_{r\mu}}({\bm\calL}_g)_{rq}. \label{eq:W}
\end{align} 
%\begin{align}
%\frac{\dd E_{\rm PHF}}{\dd {\bm\kappa}^*} =\sum_g \Biggl( \frac{\dd E_{\rm PHF}}{\dd {\bm \calW}_g} \frac{\dd {\bm \calW}_g}{\dd  {\bm\kappa}^*} + \frac{\dd E_{\rm PHF}}{\dd\tilde {\bm \calL}_g} \frac{\dd\tilde {\bm \calL}_g}{\dd  {\bm\kappa}^*} +\frac{\dd E_{\rm PHF}}{\dd\tilde {\bm \calR}_g} \frac{\dd\tilde {\bm \calR}_g}{\dd  {\bm\kappa}^*}\Biggr).
%\end{align}  
Then, the $\tilde{\bm \calL}_g$ dependence of $E_{\rm PHF}$ (or $E_g$) only appears naturally in the transition density matrix in the AO basis written as
\begin{align}
%\pgAO  = \bS^{-1} \Rg \bC_o \Woo \bC_o^\dag =  \bS^{-1} \Rg \bC_o \tilde {\bm\calL}_g^{o}
(\pg)_{\mu\nu} = (\tilde{\bm\calR}_g)_{\mu i} (\Woo)_{ij}^{-1}  (\tilde{\bm\calL}_g)_{j \mu}.
\end{align}
and there is no dependence in $n_g$, because it can be completely expressed by ${\bm\calW}_g$.
Thus, the energy-weighted density matrix for PHF gradient may be given simply by
\begin{align}
W_{pq} &= - C_{\mu p}^*\sum_g w_g n_g \frac{\dd E_g} {\dd (\pg)_{\alpha\beta}} \frac{\dd (\pg)_{\alpha\beta}}{\dd (\tilde {\bm \calL}_g)_{r\mu}} ({\bm \calL}_g)_{rq}\bre
&= - C_{\mu p}^*\sum_g w_g n_g (\Fg)_{\beta\alpha}[\bS^{-1}\Rg \bC_o]_{\alpha k}\delta_{\mu\beta} (\Lg)_{kq}\bre
%&= - \sum_g w_g n_g [\Fg \bC^\dag \Rg \bC_o {\bm\calL}_g^o]_{pq}\bre
&= -\sum_g w_g n_g (\Fg \pg)_{pq},
\end{align} 
where we have used
\begin{align}
&\frac{\dd (\pg)_{\alpha\beta}}{\dd (\tilde {\bm \calL}_g)_{k\mu}} = [ \bS^{-1} \Rg \bC_o]_{\alpha k}\delta_{\mu\beta}.\\
&\frac{\dd (\pg)_{\alpha\beta}}{\dd (\tilde {\bm \calL}_g)_{a\mu}} = 0.
\end{align}
This is consistent with the energy-weighted density matrix of PHF previously derived in Refs.[\onlinecite{Schutski14,Uejima16}],
%\begin{align}
%W_{\mu\nu} =  \sum_g w_g n_g \Bigl((E_g - E_{\rm PHF}) \rho_{\nu\mu}(g) - \rho_{\nu\lambda} F_{\lambda\sigma}\rho_{\sigma\mu}\Bigr)
%\end{align} 
%which is in the MO basis
\begin{align}
{\bf W} = \sum_g w_g n_g \Bigl( (E_g - E_{\rm PHF})\pg - \pg \Fg \pg \Bigr). \label{eq:Wgus}
\end{align} 
which, using the variational condition of the PHF Fock matrix
\begin{align}
{\bm \calF} = \frac{\dd E_{\rm PHF}}{\dd {\bm\kappa}^*} = \sum_g w_g n_g \Bigl( (E_g - E_{\rm PHF})\pg + {\bm\eta}_g \Fg \pg \Bigr) = {\bf 0}\label{eq:PHFFock},
\end{align} 
becomes
\begin{align}
 {\bf W} = - \sum_g w_g n_g \left({\bm\eta}_g \Fg \pg + \pg \Fg \pg  \right) = - \sum_g w_g n_g \Fg \pg.
\end{align}
Hence, the equivalence is proven.
\fi

\bibliographystyle{aip}
%\bibliography{refs}

\end{document}